\documentclass[twocolumn]{aastex631}

\def\sun{\hbox{$\odot$}}
\def\earth{\hbox{$\oplus$}}
\def\la{\mathrel{\hbox{\rlap{\hbox{\lower4pt\hbox{$\sim$}}}\hbox{$<$}}}}
\def\ga{\mathrel{\hbox{\rlap{\hbox{\lower4pt\hbox{$\sim$}}}\hbox{$>$}}}}

\def\deg{{^\circ}}

\def\fm{\hbox{$.\!\!^{\rm m}$}}

\def\fdg{\hbox{$.\!\!^\circ$}}

\newcommand{\Mag}{^m\llap{.\thinspace}}

\newcommand{\B}{{$B$}}
\newcommand{\V}{{$V$}}
\newcommand{\R}{{$R$}}

\newcommand{\J}{{$J$}}

\newcommand{\K}{{$K_s$}}

\def\apjl   {{ApJL }}
\def\apjs   {{ApJS }}

\def\aap    {{A\&A }}

\usepackage[para,online,flushleft]{threeparttable} 
\usepackage{booktabs} 

\begin{document}

\title{The $K_s$-band luminosity function of the rich cluster VC04 in the Vela supercluster}

\author[0000-0001-7075-0155]{N. Hatamkhani}
\affiliation{Department of Astronomy, University of Cape Town, Private Bag X3, University Ave N, Rondebosch, Cape Town, 7700, South Africa}
\affiliation{South African Astronomical Observatory, PO Box 9, 1 Observatory Road, Observatory, Cape Town, 7925, South Africa}

\author{R. C. Kraan-Korteweg}
\affiliation{Department of Astronomy, University of Cape Town, Private Bag X3, University Ave N, Rondebosch, Cape Town, 7700, South Africa}

\author{S. L. Blyth}
\affiliation{Department of Astronomy, University of Cape Town, Private Bag X3, University Ave N, Rondebosch, Cape Town, 7700, South Africa}

\author{R. E. Skelton}
\affiliation{South African Astronomical Observatory, PO Box 9, 1 Observatory Road, Observatory, Cape Town, 7925, South Africa}



\begin{abstract}
To learn more about the properties of the Vela Supercluster (VSCL) located behind the Milky Way at $cz\sim 18000$~km~s$^{-1}$, we determined the $K_s$-band Luminosity Function (LF) of VC04, the richest known galaxy cluster in the VSCL, and two other VSCL clusters (VC02 and VC08).
The galaxy sample is based on NIR observations which are complete to an extinction-corrected absolute magnitude of $M_{Ks}^o<-21\fm5$ ($\sim 2\fm5$ below $M_K^*$), within the  clustercentric radius of $r_c<1.5~$Mpc ($\sim 70 \%$ of the Abell radius). For VC04 we obtained 90 new spectroscopic redshifts of galaxies observed with the 11-m Southern African Large Telescope. We found the Schechter parameters of the  VC04 LF to be $M^*=-24\fm41\pm 0.44$, $\alpha=-1.10\pm 0.20$ and $\phi^*=8.84\pm 0.20$. Both the redshift data and the LF confirm VC04 to be a rich, not yet fully relaxed cluster.
We independently determined the LF of VC04 on membership defined by the Red-Sequence method and demonstrated that this method can be used in the absence of high spectroscopic coverage over a cluster. This allowed us to determine the LFs of VC02 and VC08. We also derived the LFs of the Coma, Norma and Virgo clusters to similar depth and extent as the VSCL clusters. We found that the Schechter parameters of VC04 are within $1\sigma$ uncertainties of these local clusters, as well as VC02 and VC08. We do not find significant differences between the LFs in the different cluster environments probed in this work down to $M_K^*+2\fm5$.


\end{abstract}



\section{Introduction} \label{sec:intro}

The Luminosity Function (hereafter LF) or space density of galaxies, $\phi$(L) is the number density of galaxies in a given luminosity range. The LF of a population of galaxies provides us with insight into the physics of  the formation and evolution of these galaxies \citep[e.g.][]{benson2003,henriques2015,rodrig2017,mitchell2018}. For instance, the impact of environment on galaxy evolution can be studied by comparing the LFs of galaxies in different environments \citep[see Fig.~1 in][]{binggeli1988}, which displays the difference between the cluster and field LFs.

\citet{abell1975} was the first to investigate the LF of clusters and determine the characteristics of the ``general'' LF integrated over all morphological types. \citet{binggeli1988} showed that the overall LF of galaxy clusters is a combination (sum) of different populations: a Gaussian function for ellipticals and early spirals, and exponential with different slopes for gas-rich dwarf galaxies like irregulars or dwarf ellipticals. \citet{binggeli1988} proposed that the LF of each morphological type (LFT) could be universal. This suggestion was supported by different authors \citep[e.g.][]{andreon1997,jerjen1997,andreon1998}. They compared  LFTs  derived  in  different  environments and found  that  the  shape  of  the  LF  of elliptical, early- and late-type galaxies  seems  to  be  independent of the environmental density. However, other authors \citep[e.g.][]{bromley1998,marinoni1999,Cuesta-Bolao2003} found that the LFTs significantly differ between environments of different densities. They proposed that, in general the LF should be stated as a combination of both density and morphology.

LFs derived at different wavelengths are sensitive to different galaxy evolution mechanisms which result in the visible morphology of galaxies seen today. 
The impact of extinction is much weaker at infrared wavelengths than in the optical ones.  Moreover, the Near-Infrared (NIR) $K$-band is sensitive to the underlying old stellar mass function and less sensitive to the star formation effects in galaxies than the optical wavelengths. As a result $K$-band LFs differ from optical ones. The $K$-band LF is also an excellent tracer of the stellar mass of a cluster which can be compared  with that predicted from theoretical models \citep{devereux1987,gavazzi1996,bell2001,bell2003}. A further advantage of working in the $K$-band is the small effect of $k$-correction; the conversion of the apparent magnitude to the equivalent measurement in its rest-frame. This makes the NIR LFs well-suited for evolutionary studies over a range of redshifts.

\citet{devereux2009} performed a detailed study on the dependence of the $K_s$-band LF on the morphological types of galaxies. 
The shape of the $K_s$-band LF itself is strongly dependent on the composition of the morphological types of galaxies.  
There is a dip in the total $K_s$-band LF of galaxies which happens roughly at $M_K - 5\,\log\,h_{70} = -21\fm0$. The dip is situated where the space density of elliptical and lenticular galaxies drops and late-type spirals and dwarf ellipticals take over \citep[see Figure~1 in ][]{devereux2009}. This dip has also been observed in other NIR LFs of clusters \citep[e.g.][]{merluzzi2010} and also at optical wavelengths. For example, \citet{mobasher2003} found a dip at $M_B\sim -18\fm0$ where the transition between giants and dwarf galaxies occurs.

The late-type spiral and irregular galaxies are the dominant population in the field. The most recent determination of the $K$-band field LF was done by \citet{jones2006}, using the 6dFGS at a median redshift of $\overline{z}=0.054$ which is complete out to $M_{K}\la -24\fm0$. Their derived Schechter parameters were $M^*_K= -24\fm60\pm0.03$ and $\alpha = -1.16\pm0.04$. However their Schechter function could not reproduce the flat faint-end slope and decline at the bright end simultaneously.

In this paper we analyze the NIR galaxy population of the clusters within the Vela Supercluster (VSCL) through their LFs. The VSCL is an extended structure across the Zone of Avoidance (ZoA) located at $cz\sim18000\,$km~s$^{-1}$ \citep[$\ell=272\fdg5\pm20\deg, b=0\deg\pm10\deg$,][]{kraan2017}, which may add an appreciable contribution to the residual bulk flow generated at distances beyond $\sim 16000$\,km\,s$^{-1}$ \citep{springob2014,springob2016,scrimgeour2016}. While the redshift coverage over the VSCL survey area \citep[described in][]{kraan2017} provides substantial evidence of  a supercluster consisting of two - potentially merging - walls,  the survey remains sparsely sampled. \citet{sorce2017} and \citet{courtois2019} present independent support of the prominence of the VSCL based on reconstructions.
A key signature of the richness of superclusters is their large number of massive 
galaxy clusters. Apart from two CIZA clusters
\citep[CIZA J0812.5-5714 and CIZA J0820.9-5704;][]{kocevski2006}  at the ZoA edges of the VSCL survey area, no clusters were previously known to substantiate this claim.   However, the VSCL redshift survey uncovered about twenty galaxy concentrations with velocity  dispersions above $\sigma \ga 400$~km~s$^{-1}$ indicative of potential galaxy clusters.
To learn more about these clusters we embarked on a NIR survey of the richest of these cluster candidates. 

Details of the NIR photometry of the galaxy clusters is presented in \citet{hatamkhani2023} (hereafter Paper~1). The completeness magnitude limit of the NIR VSCL survey is $K^o_s=15\fm5$ which translates to $M^o_{Ks} - 5\,\log\,h_{70} = -21\fm5$ at the distance of the VSCL, which is $\sim 2\fm5$ below the $M_K^*$. The current (second) paper uses the NIR data to determine the $K_s$-band LFs to learn more about the VSCL clusters and their richnesses. The LFs of the VSCL clusters are expected to be dominated by the ellipticals and early spirals with their morphological mix of galaxies depending on the type of cluster but they do not reach the dwarf regime. We particularly emphasise the VC04 cluster, because it is one of the richest known clusters in the VSCL and may well be comparable to the Coma and Norma clusters (see Paper~1). We used the 11-m Southern African Large Telescope (SALT) to obtain high spectroscopic coverage over the VC04 cluster area.  This data will allow for an in-depth study of this potentially rich cluster.

The structure of this paper is as follows. Section~\ref{sec:data} describes the NIR and optical spectroscopic data used in this paper. It also includes the catalog of 90 new spectroscopic redshifts of galaxies in VC04 observed with SALT. Different methods of deriving LFs are explained in Section~\ref{sec:lfmethod}. This is followed by the derivation of the $K_s$-band LFs of VC04 and two other VSCL clusters in Section~\ref{sec:vc04lf}. A discussion and comparisons of the VC04 LF parameters with those of other clusters with different redshifts and environments are represented in Section~\ref{sec:discussion} and a summary of the results is given in Section~\ref{sec:summary}.
Throughout this paper we use a cosmological model with parameters: $H_{0}=70\,\mathrm{km~s\textsuperscript{-1}~ Mpc\textsuperscript{-1}}, \Omega_{M}=0.3$ and $\Omega_{\Lambda}=0.7$.

\section{DATA}
\label{sec:data}
\subsection{NIR data}
The NIR ($JHK_s$) observations of six VSCL cluster candidates were conducted using the InfraRed Survey Facility (IRSF) telescope, a 1.4m telescope situated at the South African Astronomical Observatory (SAAO) site in Sutherland, during three observing runs (2015-2019). The process of observations, data reduction and creating the photometric catalogs of galaxies is explained in detail in Paper~1. We showed there that five out of the six cluster candidates have the characteristics of a galaxy cluster while one has a more filament-like structure, and hence unlikely to be a cluster. The extinction-corrected completeness magnitude limit for the NIR survey is $K_s^o<15\fm5$  which is $\sim 2\fm0$ deeper than the 2MASS Extended Source catalog \citep[2MASX,][]{jarrett2000A}. All cluster data are complete to a clustercentric radius of 1.5~Mpc. We use the  photometric catalogs of five clusters (VC02, VC04, VC05, VC08 and VC11) to measure their LFs up to the completeness magnitude limit and radius of the survey. \citet{said2015} demonstrated that isophotal magnitudes are preferable to total magnitudes when working with galaxies in the ZoA. Our LFs and completeness derivations are therefore based on isophotal photometry ($K_{s20}$).

\subsection{Existing spectroscopy}
\citet{kraan2017} used optical \citep[$B\la 18\fm5$,][]{kraan2000catalog} and 2MASX galaxies \citep[$K_s\la 13\fm5$,][]{skrutskie2006} to identify potential galaxy clusters in the VSCL. Their spectroscopic observations of the VSCL galaxies were conducted with AAOmega+2dF on the 4-m Australian Telescope, and complemented by spectroscopy with SALT in the dense cores of potential galaxy clusters in the VSCL. 
The resulting redshift distribution on which the discovery of the VSCL was based is shown in Figs.~2 and 3 of \citet{kraan2017}.

In Paper~1 we performed a deep NIR study on the six observed VSCL cluster candidates and identified their constituent galaxies. All galaxies with initial redshifts from the \citet{kraan2017} work were recovered in our NIR imaging and have IRSF photometry confirming that we are sensitive to both early- and late-type galaxies. The NIR VSCL data are complete to $K_s^o<15\fm5$ ($M_{Ks}^o<-21\fm5$), and therefore most of the fainter galaxies ($K_s^o\ga 13\fm0$, $M_{Ks}^o\ga -24\fm0$) which were not previously identified do not have redshifts. The fraction of galaxies with spectroscopic data varies in each of the observed VSCL clusters (see Table~13 in Paper~1). 
Our NIR study showed that VC04 is a the richest cluster amongst the five and is comparable to the Norma cluster. Therefore, we focused our first detailed analysis on VC04. To enable robust determination of the VC04 LF we needed to identify its cluster members. We therefore, followed up with a campaign on SALT to obtain spectroscopic redshifts of all the galaxies at the core and many at the outskirts of VC04.

\subsection{New SALT spectroscopy}
\subsubsection{Observations and data reduction} 

Situated at the SAAO Sutherland site, SALT is the largest single optical telescope in the southern hemisphere and among the largest in the world, that is principally designed for spectroscopy \citep{buckley2006}. 
In 2019 we put forward a multi-semester SALT proposal (2019-2-SCI-026, PI: Hatamkhani) to observe galaxies in the rich VC04 cluster, with the Robert Stobie Spectrograph \citep[RSS,][]{burgh2003}, in Multi-Object Spectroscopy (MOS) mode. The latter uses custom designed slit masks. SALT is ideally suited for these follow-up spectroscopic observations of faint NIR galaxies, because its circular FOV of $8'$ is practically the same as the IRSF FOV  ($7.7'$), with which we originally observed the clusters. As such, we could set up MOS fields in the same configuration as for the IRSF mosaics of the clusters. Each slitlet in a mask was assigned to a target galaxy selected from the NIR photometric catalog of the galaxies. We set the slitlets to have a width and length of $1.5'' \times 10''$, to ensure that sky subtraction could be done accurately given the seeing conditions at Sutherland \citep[median value of FWHM=$1.3''$,][]{catala2013}. The magnitude range of the targeted galaxies is $11\fm0<K_{s}<17\fm0$ which is equivalent to $15\fm0 \lesssim B \lesssim 21\fm0$ assuming $B-K_{s}=3\fm92$ for elliptical galaxies \citep{jarrett2000b-k}. This value is appropriate for elliptical galaxies which are typically found in the higher density regions of galaxy clusters \citep[morphology-density relation,][]{dressler1980}. 
We were able to target 10–16 galaxies in each cluster field per slit mask, using a script written by David Gilbank (private communication).  We prioritized galaxies with magnitudes below $B\lesssim 19\fm5$ which corresponds to our extinction-corrected magnitude completeness limit of $K_{s}^o<15\fm5$. The position angle of each mask was chosen to optimise the number of galaxies in each mask. In total, we targeted 249 galaxies in VC04 distributed in 19 masks, of which 13 were observed.

To obtain redshifts for the selected galaxies, we optimised the wavelength range to include the Ca H \& K lines, and the 4000~\AA~ break for early type galaxies and Balmer and OIII emission lines in late type galaxies. These fall between 3800 to 6800~\AA~ in the rest frame, corresponding to the observed wavelengths of 4100 to 7000~\AA~ at the redshift of VC04. We used the PG0900 Volume Phase Holographic grating, with camera station 29\fdg5 and grating angle 14\fdg75.
Initially we requested an uninterrupted exposure time of 1200~s per mask. This led to 2400~s per mask in total, including the overhead for MOS acquisition of 900~s, and the time needed for calibration (arc and flats). Standard flats and arc calibrations were taken after each science observation.  
After examination we found that the spectra of galaxies that were obtained in the first semester had low signal-to-noise ratios (SNRs). We therefore, increased the exposure time to $1800~$s. 
We selected 4–8 bright stars per cluster mask from the USNO-A2.0 catalog \citep{monet1998} for aligning the slit masks during acquisition. Because the spectroscopic observations were taken in grey time, the reference stars were selected to be bright, $11\fm0 - 15\fm0$ in the $r$ band. This made the acquisition of the reference stars quick (within 3~s).

\begin{figure}
\begin{center}
    \includegraphics[width=\columnwidth, height=\textheight, keepaspectratio]{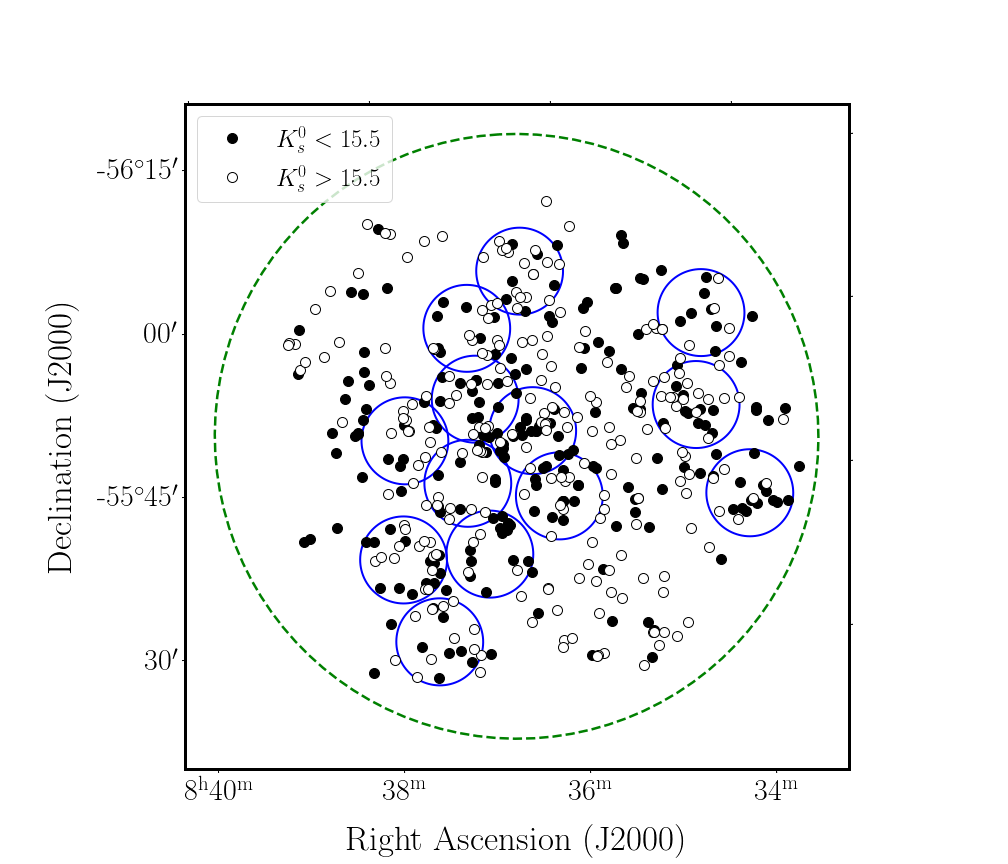}
    \caption{The distribution of galaxies in the VC04 region. The filled circles denote galaxies up to completeness magnitude limit of $K^o_{s}<15\fm5$, while the open circles are fainter galaxies. The larger blue circles show the SALT mask locations. The large green dashed circle shows the Abell radius.} 
    \label{fig:saltmasks}
    \end{center}
\end{figure}

In total 13 high priority masks were observed during three semesters (2019-2021). Fig.~\ref{fig:saltmasks} shows the positions of the observed masks in VC04. The aim was to first cover the center of the cluster followed by other high density regions in the surrounding areas. The figure also displays the distribution of the galaxies within the Abell radius of VC04, color coded based on their magnitudes. Table~\ref{tab:masksprop} lists the properties of the observed masks in VC04 and enumerates the number of selected galaxies in each mask based on their magnitudes in $K_{s}$ band. The first priority per mask were galaxies within the completeness magnitude limit of $K_{s}^o<15\fm5$ (N1), then galaxies with $K_{s}^o>15\fm5$ (N2), and finally galaxies with existing redshift measurements \citep{kraan2017}, which are limited to our NIR brightest galaxies ($K_s^o<13\fm0$), and are useful for comparisons with the resulting SALT redshifts (N3).

\begin{table}
\caption{Properties of the observed masks in the VC04 cluster.}
\footnotesize
\hspace*{-1cm}
\centering
\begin{threeparttable}
\begin{tabular}{ccccc}
\toprule
\toprule
 Field name & RA, Dec & N1\tnote{a}  & N2\tnote{b} & N3\tnote{c} \\
  & (deg, deg)&  &  &  \\ 
 \midrule
 VC04F-1-1&129.038, -55.723 & 8&7 &3 \\
 VC04F-1-3 &128.520, -55.710 &7 &5 & \\ 
 VC04F-1+1 &129.285, -55.750 &5 &8 &1 \\
 VC04F-2+1 &129.236, -55.640 &9 &4 &2 \\
 VC04F-2+2 &129.471, -55.638 &6 &10 &3 \\ 
 VC04F-3+2 &129.385, -55.510 &4 &8 & \\ 
 VC04F+1-1 &129.100, -55.825 &10 &6 &2 \\
 VC04F+1-3 &128.650, -55.850 &6 &10 & \\ 
 VC04F+1+1 &129.252, -55.878 &7 &7 &2 \\
 VC04F+1+2 &129.450, -55.820 &4 &9 & \\ 
 VC04F+2-3 &128.620, -55.990 &5 &5 & \\ 
 VC04F+2+1 &129.264, -55.987 &6 &7 &1 \\
 VC04F+3+1 &129.110, -56.070 &3 &12 &1 \\

 \bottomrule
\end{tabular}
\begin{tablenotes}
\item[a] Number of galaxies with $K_{s}^o<$ 15\fm 5.
\item[b] Number of galaxies with $K_{s}^o>$ 15\fm 5.\\
\item[c] Number of galaxies with known redshifts from previous spectroscopy surveys \citep{kraan2017}, which are brighter galaxies with ($K_s^o<13\fm0$).
\end{tablenotes}
\end{threeparttable}
\label{tab:masksprop}
\end{table}

We reduced the SALT data using the fully automated RSSMOSPipeline\footnote{
\url{https://github.com/mattyowl/RSSMOSPipeline}}, that is written in pure Python and performs flat field corrections, wavelength calibration, extraction and stacking of 1D-spectra \citep[described in ][]{hilton2018}.

\subsubsection{Redshift measurements and results}
The galaxy redshifts were derived using the \texttt{RVSAO/XCSAO} package in \texttt{IRAF} \citep{kurtz1998}. They were verified with the RSSMOSPipeline visualisation tool which displays the obtained spectra of galaxies with SDSS spectral templates\footnote{\url{http://classic.sdss.org/dr5/algorithms/spectemplates/}} corresponding to different galaxy morphologies overlaid.   

\begin{figure}
\begin{center}   
    \includegraphics[width=\columnwidth, height=0.4\textheight, keepaspectratio]{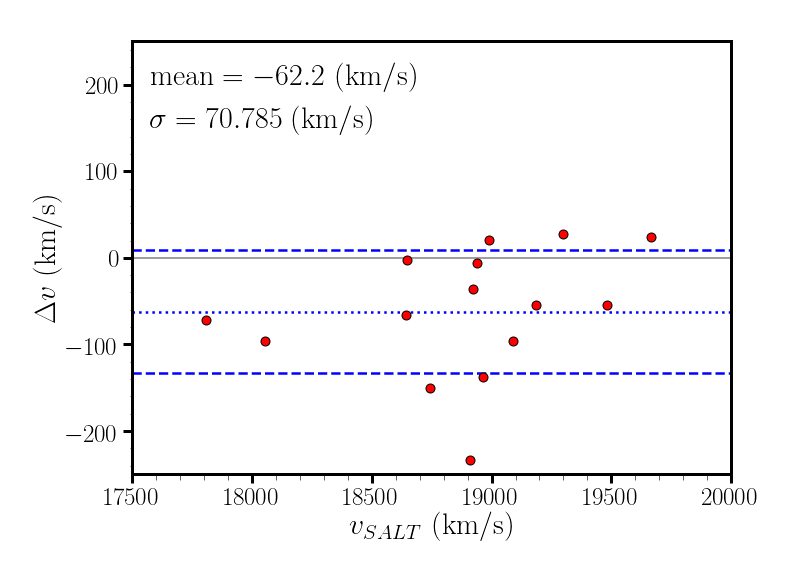}
    \caption{Comparison of the recession velocities for galaxies re-observed with SALT. The dotted line indicates the mean value of $\Delta v$ and the dashed lines indicate the $1\sigma$ dispersion around the mean. The mean and standard deviation are given in the top left corner of the plot.} 
    \label{fig:deltaz}
    \end{center}
\end{figure}

Based on the SALT observations we obtained the spectra of 175 galaxies from the 13 observed masks in VC04, of which 90 had high enough SNRs to yield a reliable redshift. The velocity catalog of the 90 galaxies is given in Table.~\ref{tab:redshift-catalog}, which lists the coordinates of galaxies, their $JHK_s$ isophotal magnitudes and their heliocentric velocities with corresponding errors. 
From previous spectroscopic surveys \citep{kraan2017}, we had redshifts of $88$ galaxies distributed over the Abell radius of VC04, of which 15 galaxies were re-observed with SALT. We compared the SALT redshifts with the redshifts from the \citet{kraan2017} work as a consistency check, by computing the velocity difference: 
$$\Delta v=v_{\mathrm{SALT}}-v_{\mathrm{previous~survey}}.$$ 
The result of the velocity comparison is shown in Fig.~\ref{fig:deltaz}. The values of the mean recession velocity difference and the standard deviation are displayed in the top left corner of Fig.~\ref{fig:deltaz} ($\langle \Delta v \rangle = -62.200\pm 70.785$~km~s$^{-1}$). The values of $\langle \Delta v \rangle$ are consistent with zero within the $1\sigma$ uncertainties.

\startlongtable
\begin{deluxetable}{ccccccc}
\tabletypesize{\footnotesize}
\tablecaption{Velocity catalog of galaxies in the VC04 cluster which includes 90 spectra observed with SALT. $v\pm\Delta v$ shows the heliocentric velocity with its corresponding error. $J_{20}$, $H_{20}$, $K_{s20}$ are the  $K_{s20}$ fiducial isophotal magnitudes. The errors on the magnitudes are given in the full NIR catalog presented in Paper~1.} \label{tab:redshift-catalog}

\startdata 
ID& RA & Dec & $J_{20}$ & $H_{20}$& $K_{s20}$ & $v\pm\Delta v$ \\
 & deg & deg& mag & mag & mag & km s$^{-1}$ \\ 
 \midrule
1&129.077& $-55.766$& 15.624&14.909&14.532&$17990 \pm 27$\\
2&129.102&$-55.750$&15.475&14.654&14.208&$19667\pm18$\\
3&129.024&$-55.762$&16.391&15.672&15.232&$19120\pm72$\\
4&129.098&$-55.741$&17.155&16.527&15.474&$18929\pm58$\\
5&129.031&$-55.752$&17.942&17.192&16.463&$28306\pm 96$\\
6&129.132&$-55.728$&18.239&17.204&16.322&$53004\pm66$\\
7&129.021&$-55.744$&17.190&16.469&15.932&$20739\pm 88$\\
8&128.985&$-55.737$&16.232&15.638&14.725&$17078\pm 93$\\
9&129.029&$-55.714$&14.575&13.720&13.220&$18205\pm 15$\\
10&128.998&$-55.713$&16.641&15.899&15.552&$18627\pm 60$\\
11&129.031&$-55.703$&17.061&16.355&15.776&$17806\pm 24$\\
12&129.060&$-55.691$&16.875&15.975&15.402&$27359\pm 80$\\
13&129.031&$-55.685$&13.692&12.953&12.573&$18910\pm 60$\\
14&128.517&$-55.698$&16.145&15.440&14.929&$33527\pm 49$\\
15&129.364&$-55.764$&15.550&14.920&14.633&$19099\pm 59$\\
16&129.211&$-55.753$&16.176&15.372&14.859&$44592\pm 35$\\
17&129.210&$-55.749$&16.033&15.27&15.024&$18155\pm 75$\\
18&129.311&$-55.711$&16.391&15.567&15.225&$18878\pm 70$\\
19&129.242&$-55.704$&17.545&16.65&15.876&$19177\pm 61$\\
20&129.196&$-55.697$&15.780&15.044&14.647& $26289\pm 69$\\
21&129.259&$-55.671$&16.838&16.148&15.840& $19754\pm 29$\\
22&129.203&$-55.679$&15.869&15.123&14.810&$18549\pm 53$\\
23&129.175&$-55.682$&16.125&15.414&15.052&$18516\pm 40$\\
24&129.200&$-55.671$&15.321&14.581&14.174& $18720\pm 23$\\
25&129.288&$-55.631$&16.009&15.297&14.919& $18850\pm 40$\\
26&129.175&$-55.629$&16.203&15.432&15.060& $19113\pm 18$\\
27&129.132&$-55.625$&14.524&13.843&13.418& $18056\pm 25$\\
28&129.251&$-55.581$&14.998&14.300&13.875& $18743\pm 56$\\
29&129.298&$-55.615$&18.010&17.284&16.961&$32769\pm 52$\\
30&129.464&$-55.684$&17.254&16.487&15.719&$20980\pm 56$\\
31&129.397&$-55.663$&17.211&16.407&15.784&$22281\pm 42$\\
32&129.503&$-55.686$&15.665&14.947&14.550&$19606\pm 16$\\
33&129.388&$-55.630$&16.245&15.535&15.185&$18417\pm 47$\\
34&129.547&$-55.668$&14.175&13.408&13.020&$19300\pm 25$\\
35&129.395&$-55.620$&16.891&16.143&15.823&$18098\pm 37$\\
36&129.496&$-55.642$&16.949&16.202&15.825&$18250\pm 29$\\
37&129.532&$-55.643$&17.288&16.535&16.304&$25431\pm 51$\\
38&129.486&$-55.595$&14.334&13.590&13.220&$19089\pm 21$\\
39&129.371&$-55.564$&15.770&16.727&15.786&$54181\pm 96$\\
40&129.399&$-55.562$&nan&14.584&14.167&$17809\pm 18$\\
41&129.289&$-55.526$&17.133&16.654&16.092&$33126\pm 76$\\
42&129.434&$-55.503$&15.767&14.922&14.440& $18474\pm 33$\\
43&129.452&$-55.458$&17.867&16.825&16.105&$33088\pm 49$\\
44&129.012&$-55.850$&17.488&16.547&15.844&$24847\pm 50$\\
45&129.116&$-55.842$&15.655&14.927&14.605& $17805\pm 50$\\
46&129.066&$-55.833$&17.130&16.432&16.194&$19362\pm 61$\\
47&129.078&$-55.828$&15.789&15.027&14.720&$18545\pm 19$\\
48&129.092&$-55.824$&15.866&15.129&14.799&$18961\pm 54$\\
49&129.129&$-55.819$&14.492&13.749&13.345&$18644\pm 19$\\
50&129.183&$-55.806$&13.533&12.772&12.401&$19284\pm 25$\\
51&129.121&$-55.800$&16.820&16.100&15.703& $18648\pm 8$\\
52&129.192&$-55.796$&15.505&14.752&14.356&$18922\pm 49$\\
53&129.181&$-55.787$&15.754&14.924&14.557& $18942\pm 22$\\
54&129.077&$-55.766$&15.624&14.909&14.532& $18113\pm 36$\\
55&128.731&$-55.895$&17.732&16.913&16.015&$34731\pm 54$\\
56&128.691&$-55.899$&17.029&16.338&15.760&$35858\pm 48$\\
57&128.684&$-55.865$&15.817&15.152&14.755&$18434\pm 21$\\
58&128.643&$-55.867$&16.756&16.072&15.940&$18543\pm 21$\\
59&128.744&$-55.824$&15.895&15.203&14.901& $19516\pm 25$\\
60&128.623&$-55.797$&16.512&15.804&15.617&$19128\pm 34$\\
61&129.318&$-55.915$&17.093&16.457&16.180&$18821\pm 34$\\
62&129.261&$-55.900$&17.144&16.625&16.278&$20413\pm 2$\\
63&129.348&$-55.878$&15.342&14.700&14.294&$18939\pm 55$\\
64&129.324&$-55.873$&17.401&16.724&16.479&$46160\pm 37$\\
65&129.140&$-55.883$&15.744&15.064&14.750&$19482\pm 50$\\
66&129.264&$-55.849$&16.188&15.493&15.169&$18357\pm 32$\\
67&129.230&$-55.832$&18.278&17.455&16.805&$28920\pm 59$\\
68&129.235&$-55.822$&15.564&14.847&14.470&$19185\pm 21$\\
69&129.220&$-55.818$&16.489&15.773&15.318&$25053\pm 38$\\
70&129.451&$-55.865$&17.660&17.026&16.476&$30751\pm 75$\\
71&129.450&$-55.843$&16.509&15.849&15.369& $19130\pm 8$\\
72&129.370&$-55.841$&16.596&15.902&15.366&$14491\pm 12$\\
73&129.439&$-55.834$&18.465&18.547&16.594&$19949\pm 87$\\
74&129.498&$-55.792$&15.720&15.015&14.604&$18648\pm 20$\\
75&129.459&$-55.793$&16.791&15.952&15.380&$22554\pm 5$\\
76&128.647&$-55.990$&16.714&16.039&15.468& $18598\pm 12$\\
77&128.730&$-55.968$&16.352&15.613&15.239&$19029\pm 68$\\
78&129.192&$-56.020$&17.630&16.892&16.124&$33201\pm 44$\\
79&129.262&$-56.020$&16.156&15.388&15.270& $18867\pm 42$\\
80&129.324&$-56.028$&16.183&15.453&15.149&$18882\pm 40$\\
81&129.251&$-55.969$&16.883&16.123&16.032&$20858\pm 85$\\
82&129.190&$-55.946$&14.969&14.204&13.799&$18458\pm 24$\\
83&129.345&$-55.958$&13.799&13.125&12.757&$8269\pm 27$\\
84&129.162&$-56.118$&17.394&16.665&16.174&$32392\pm 86$\\
85&129.127&$-56.112$&16.561&15.765&15.456&$19245\pm 16$\\
86&129.131&$-56.055$&16.423&16.232&15.439&$19015\pm 61$\\
87&129.175&$-56.022$&16.710&15.958&15.749&$18478\pm 46$\\
88&129.123&$-56.014$&17.380&16.680&16.122&$17210\pm 20$\\
89&129.102&$-56.008$&13.459&12.703&12.315&$18966\pm 33$\\
90&129.343&$-56.008$&15.460&14.603&14.105&$18990\pm 24$\\
\enddata

\end{deluxetable}

\begin{figure}
\begin{center}   
    \includegraphics[width=\columnwidth, height=\textheight, keepaspectratio]{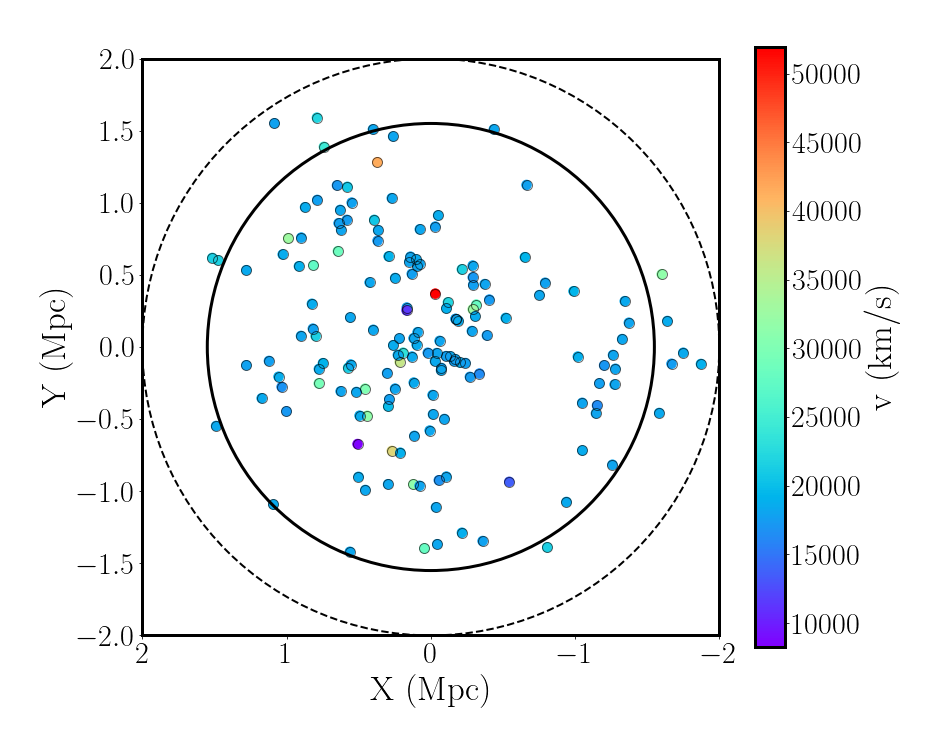}
    \caption{The distribution of 162 galaxies with available redshift values in the VC04 region. The color bar on the right represents the recession velocities of galaxies. The dashed and solid black circles show the Abell and the completeness radii respectively.} 
    \label{fig:vc04redshift}
    \end{center}
\end{figure}    

\begin{figure}
\begin{center}   
    \includegraphics[width=0.9\columnwidth, height=\textheight, keepaspectratio]{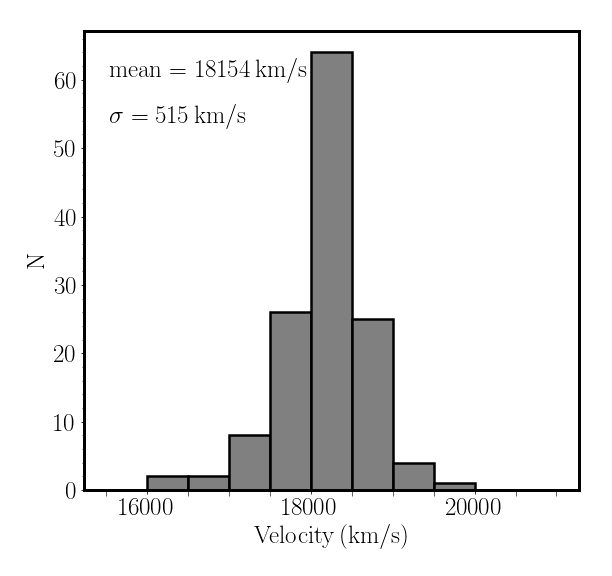}
    \caption{Recession velocity histogram of VC04 for 132 galaxies with spectroscopic redshifts after removing  foreground and background galaxies by iterating the $3\sigma$-clipping technique.} 
    \label{fig:vc04velohist}
    \end{center}
\end{figure}  

The combined tally of unique redshift values for galaxies in VC04 is 162, of which 127 have $K^o_s<15\fm5$. The spatial distribution of these galaxies with their recession velocities are shown in Fig.~\ref{fig:vc04redshift}. The Abell radius and the completeness radius of the NIR survey (2.1 and 1.5~Mpc respectively) are also depicted in Fig.~\ref{fig:vc04redshift}. Using the new spectroscopic data, we revised the preliminary mean recession velocity and dispersion of VC04 of $cz\sim 18196$~km~s$^{-1}$ and  $\sigma=455$~km~s$^{-1}$ (see Paper~1).
The new velocity distribution of galaxies with available redshift data in VC04 was fitted with a Gaussian and foreground and background galaxies were removed by iterating the $3\sigma$-clipping technique \citep{yahil1977}. The result is shown in Fig.~\ref{fig:vc04velohist}. Based on the velocities of the cluster member galaxies, VC04 is located at a redshift distance of $cz=18154$~km s\textsuperscript{-1}, very close to the original value and the approximate redshift distance of the primary VSCL wall at $cz\sim 18000$~km~s$^{-1}$, with a slightly higher velocity dispersion of $\sigma=515\,$km s\textsuperscript{-1}. The higher velocity dispersion of VC04 is more consistent with the richness of the cluster determined in the IRSF analysis in Paper~1.

In total there are 113 galaxies with  redshift data in the VC04 region within the completeness magnitude limit of $K_s^o<15\fm5$ and clustercentric radius of $r<1.5$~Mpc. These redshift values are key for a robust determination of the $K_s$-band LF of this rich cluster (see Section~\ref{sec:vc04lf}). 

\section{Deriving Luminosity functions}
\label{sec:lfmethod}
Before we derive the LF of VC04 and the other VSCL galaxy clusters with deep NIR photometry, we provide an overview of LFs and the suitable method to derive them for the VSCL clusters.

In 1976,  an analytic approximation to the LFs of galaxies was suggested by Paul Schechter \citep{schechter1976} that in terms of luminosity has the form
\begin{equation}
\label{eq:schech_fnl}
\phi(L) dL = \frac{\phi^* }{L^*} \left (\frac{L}{L^*} \right)^\alpha e^{- \left (\frac{L}{L^*} \right)} dL,
\end{equation}
where $\phi(L) dL$ is the number of galaxies per unit volume in the luminosity interval from $L$ to $L + dL$. The $\phi^*$ denotes number per unit volume, $L^*$ is the ``characteristic luminosity'', at which a rapid change in the slope of the LF is observed,  and $\alpha$ is a dimensionless parameter that gives the faint end slope of the LF.
In terms of absolute magnitude, equation \ref{eq:schech_fnl} has the form  

\begin{equation}\label{eq:schech_fn}
  \phi(M) dM = \phi^* 10^{0.4(\alpha + 1)(M^* - M)}e^{{-10^{0.4(M^* - M)}}} dM,
  \end{equation}
which indicates a double exponential profile.

In this paper we apply a nonlinear least-squares Marquardt-Levenberg algorithm to fit a Schechter function to the VSCL cluster LFs to derive their  characteristic magnitude, $M^*$, faint-end slope, $\alpha$, and normalisation, $\phi^*$.

\subsection{Derivation methods}
\label{subsec:der-meth}
The traditional way to compute the cluster LFs is to statistically subtract counts from a control field in the general direction of the cluster for each magnitude bin. 
To yield reliable results, the photometric calibration of the galaxies in both the cluster and the control field must be accurate. Also the control field should not be too close to or too far from the cluster. If it is too close, its counts might be contaminated by the cluster counts and if too far, it would not take into account the field nonuniformities on the scale of the angular size of the cluster \citep{paolillo2001}. Furthermore, the field contamination correction method could result in significant systematic errors \citep[for a detailed discussion see e.g.][]{andreon2005}.
Given that there exist no control field samples in the surroundings of the VSCL clusters that are deep enough in $K_s$-band to correct the IRSF VSCL clusters sample, we cannot apply this method to derive the LFs of the VSCL clusters. We therefore apply other methods which are described in the following sections.

\subsubsection{Spectroscopic membership derivation}
\label{subsubsec:spec-mem}
\citet{biviano1995} corroborated the membership fraction obtained from the color-magnitude plane by computing a statistical estimate of the same. A similar statistical approach was applied by \citet{mobasher2003} on spectroscopic data and later by \citet{skelton2009} to measure the LFs of the Coma and Norma clusters respectively.
In this method, the foreground and background contamination can be determined when a large enough fraction of galaxies within a cluster radius have spectroscopic information. 
\citet{skelton2009} compared the result of this method with the statistical field correction method and confirmed that they are in good agreement. Here, we follow the formalism described in \citet{skelton2009} to derive the LF of VC04.

\subsubsection{The Red-Sequence method}
\label{subsec:RSmethod}
The spectroscopic membership is a reliable LF determination method only when a high percentage of galaxies in a cluster have measured redshifts. Amongst the VSCL clusters only VC04 has good spectroscopic data coverage. We explored the Red-Sequence (RS) method as another alternative for the derivation of the VSCL cluster LFs when the majority of galaxies lack spectroscopic information. We will first apply the RS method on VC04 to show the consistency between the spectroscopic membership and RS methods (Section~\ref{subsec:methodcom}) with the ultimate goal to assess whether the RS method provides a reliable alternative to derive the LFs of the VSCL clusters with poor spectroscopic coverage.

In the RS method, the tight relation between the color and magnitudes in the Color-Magnitude Diagram (CMD) of early-type galaxies (dominant population in clusters) in different wavelengths regimes (including NIR), is used to identify the cluster members \citep[e.g.][]{Yee1999A,Gladders2000,papovich2010,rykoff2016}. The RS appears to be universal and has been observed even in the most distant clusters \citep[e.g.][]{bower1992,stanford1998,blakeslee2003,lopez2004,mei2006a,mei2006b,stott2009,snyder2012,strazzullo2016A}.

\section{Cluster luminosity functions in the VSCL}
\label{sec:vc04lf}

\subsection{VC04}
In this section we derive the LF of VC04 out to the clustercentric radius of $r_c<1.5~$Mpc. This extends into the outer region of the cluster, where more late-type galaxies, possibly infalling galaxies or galaxy groups are found compared to the core that is dominated by early-type galaxies. We use both the spectroscopic membership and  RS methods to investigate the consistency of the resulting LFs.

\subsubsection{Using the spectroscopic membership method}
\label{subsub:smm}
VC04 has 190 galaxies within $r_c<1.5$\,Mpc with $M_{Ks}^o<-21\fm5$ and 113 of them have redshift information (59\%). The sample of 113 galaxies consists of 2MASX and optical galaxies from previous surveys \citep{kraan2017} and IRSF galaxies which are distributed over the Abell radius of VC04. The selected 2MASX galaxies observed with AAOmega+2dF include bright ($K_s<13\fm5$), high surface brightness ellipticals and bulge-dominated spirals. The IRSF galaxies observed with SALT include fainter galaxies ($K^o_s<15\fm5$) and $\sim30$\% of them show emission lines in their spectra indicative of star forming galaxies. Therefore, our sample of galaxies in the spectroscopic membership method includes a combination of  elliptical, early-type spiral and star forming galaxies. Note, however, that \citet{devereux2009} showed, that ellipticals dominate the bright end of the total $K$-band LFs, followed by lenticulars and and bulge dominated spirals up to Sbc, it is only at the faint end ($M_{K}^o>-21\fm0$) that  later type spirals and irregulars take over. Therefore, our spectroscopic survey is not sensitive to this population at the distance range of VSCL.

\begin{figure}
\centering
        \includegraphics[width=\linewidth]{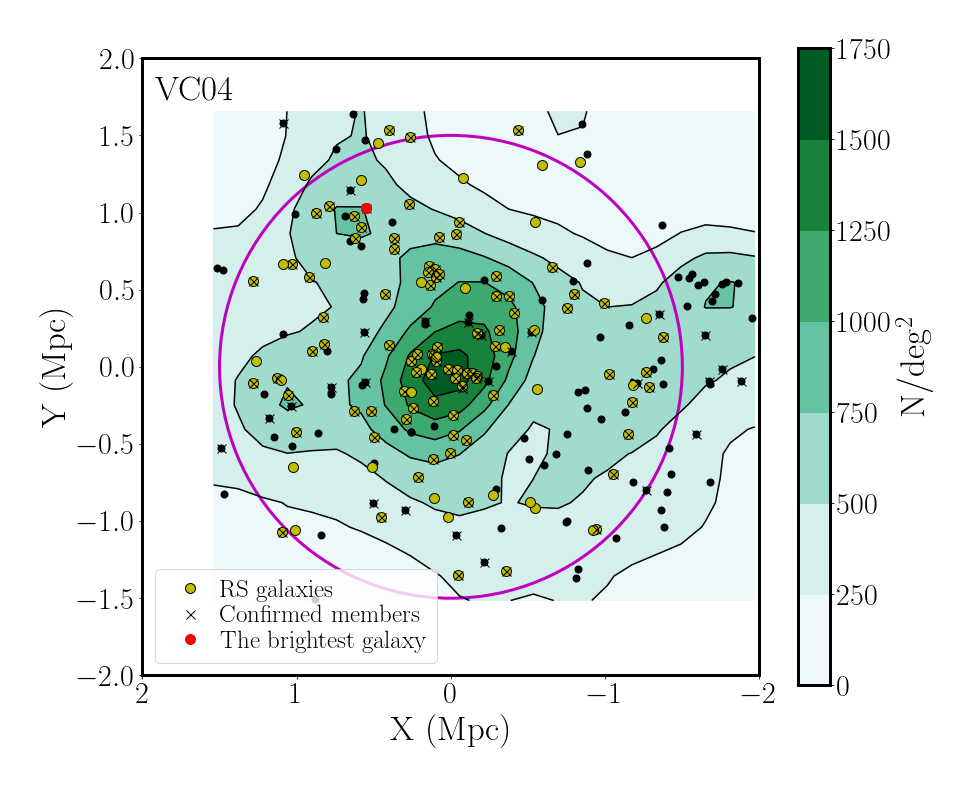} 
\caption{\label{fig:rsvc04-2} The iso-density contour map of VC04 complete to $M_{Ks}^o<-21\fm5$. The galaxies are shown as black dots and the RS galaxies as yellow dots. The spectroscopically confirmed members of VC04 are depicted as black crosses. The magenta circle represents $r_c=1.5~$Mpc.
}
\end{figure}

\begin{figure}
\begin{center}   
    \includegraphics[width=0.9\columnwidth, height=\textheight, keepaspectratio]{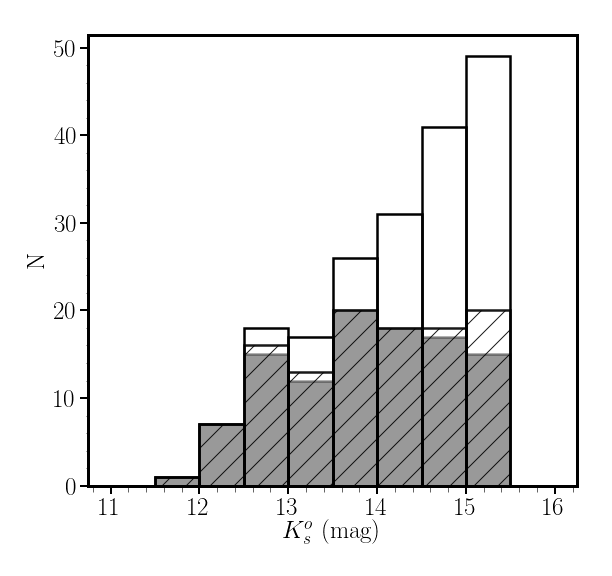}
    \caption{Distribution of the extinction-corrected $K_s^o$ magnitudes for the 190 identified galaxies in VC04. The 113 galaxies with redshift data are shown as hatched areas. The grey areas indicate cluster members (galaxies with $16000<cz<20000$~km~s$^{-1}$).} 
    \label{fig:vc04histredshift}
    \end{center}
\end{figure}

Galaxies with $16000<cz<20000$~km~s$^{-1}$ are assumed to be cluster members. There are 3 foreground and 5 background galaxies in the VC04 spectroscopic sample. Fig.~\ref{fig:rsvc04-2} shows the distribution of the spectroscopically confirmed members of VC04 in its iso-density contour map. 
The total and spectroscopic number counts in $K_s$-band magnitude intervals, used in the spectroscopic membership method to derive the VC04 LF, are shown in Table~\ref{tab:memberlistvc04} for data out to $r_c<1.5$\,Mpc. 
The distribution of the extinction-corrected $K_s$ magnitudes for all identified galaxies in the region of VC04 is shown in  Fig.~\ref{fig:vc04histredshift}. The 113 galaxies with redshift data are shown as hatched areas; the grey areas depict the distribution of cluster members (galaxies with $16000<cz<20000$~km~s$^{-1}$). Using this information we derive the LF of VC04 based on the spectroscopic membership method (described in Section.~\ref{subsubsec:spec-mem}) and list the results in Table~\ref{tab:lfspropsvc04}.

\begin{table}
\caption{\label{tab:memberlistvc04} The total and spectroscopic number counts in $K_s^o$-band magnitude intervals of the VC04 cluster out to $r_c<1.5$\,Mpc. $N_p(m)$ is the total number of galaxies per magnitude interval and $N_s(m)$ is the number with spectroscopic data, while $N_m(m)$ is the number of cluster members. $f(m)$ is the fraction of galaxies in each magnitude bin that satisfies the membership criterion.}
\hspace*{-0.5cm}
\begin{tabular}{lcccc}
\hline
\multicolumn{1}{c}{$K_s^o$ (mag)} &
\multicolumn{1}{c}{$N_p(m)$}&
\multicolumn{1}{c}{$N_s(m)$} & \multicolumn{1}{c}{$N_{m}(m)$} &
\multicolumn{1}{c}{$f(m)$} \\
\hline
11.75 & 1 & 1 & 1 & 1.00 \\ 
12.25 & 7 & 7 & 7 & 1.00 \\ 
12.75 & 18 & 16 & 15 & 0.94 \\ 
13.25 & 17 & 13 & 12 & 0.92 \\ 
13.75 & 26 & 20 & 20 & 1.00 \\ 
14.25 & 31 & 18 & 18 & 1.00\\ 
14.75 & 41 & 18 & 17 & 0.95\\ 
15.25 & 49 & 20 & 15 & 0.75 \\ 
\hline
\end{tabular}
\end{table}

\subsubsection{Using the red-sequence method}
\label{subsec:rsgal}
To find the RS galaxies, we determined an initial fit by visual inspection of the region of the CMD containing the RS relation. The points within $\pm 3 \sigma$ of this fit were kept for the next step. An iterative linear regression fit of the form  $J-K = $$\beta$$ K + c$ was then performed on the points within $3 \sigma$ of the initial fit, with $\beta$ as the slope of the RS. Galaxies within $\pm 3 \sigma$ of the final fit are the RS galaxies which were used to derive the LF of the clusters.  Errors are assumed to be Poissonian when the RS method was applied to derive the LFs. 

It should be noted that the CMD was not field corrected. \citet{stott2009} performed a statistical field correction test to a sub-sample of their low-redshift clusters and found that the slope of the field-corrected sequence varies randomly by less than $1 \sigma$ compared to the cluster sample obtained for the uncorrected sequence. The RS method can therefore, be performed on low-redshift clusters without a field correction. 

\begin{figure}
\centering
        \includegraphics[width=0.95\linewidth]{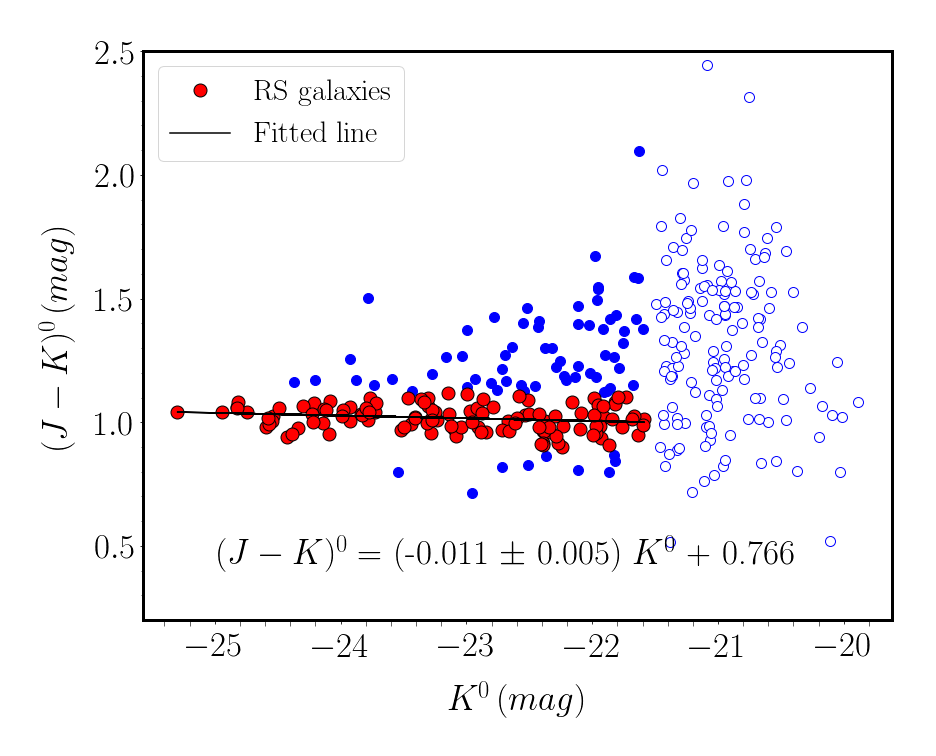} 

\caption{\label{fig:rsvc04}The RS of the VC04 cluster performed on galaxies within a clustercentric distance of $r_c=1.5$. Galaxies with $M_{K_s}^o<-21\fm5$ are shown as filled blue circles, galaxies with $M_{K_s}^o>-21\fm5$ as open circles. The RS galaxies are displayed as red circles and the RS fit is shown as black line. 
}
\end{figure}

Fig.~\ref{fig:rsvc04} shows the CMD of the VC04 cluster with the resulting RS. At the faint end the number of galaxies increases towards redder colors due to an increase in background contamination by background galaxies.  Also note that galaxy clusters at different redshifts are found to have different slopes \citep{gladders1998,lopez2004}. The RS fit of VC04 as determined from the galaxy sample delimited by $M_{Ks}^o<-21\fm5$ has the relation, $$(J-K)^o=(-0.011\pm 0.005)K^o+0.766.$$ The slope of the regression is within $1\sigma$ of the NIR RS slope found by \citet{stott2009} for the Coma cluster, i.e. $\beta =-0.017\pm 0.009$. Out of 190 galaxies within $r_c=1.5~$Mpc, 111 are identified as RS galaxies up to $M_{Ks}^o<-21\fm5$, which is very close to the number of galaxies in the spectroscopic sample. Fig.~\ref{fig:rsvc04-2} depicts the distribution of the RS galaxies in the iso-density contour map of VC04.  We use this RS sample to derive the LF of VC04 and list the LF parameters in Table~\ref{tab:lfspropsvc04}. Note that, although RS is dominated by elliptical and early-type spiral galaxies particularly at its bright end, it also includes late-type spiral and irregular galaxies at its faint end \citep[see Fig. 8 in][]{ramatsoku2020}. However, as mentioned before, we are sensitive to normal spiral galaxies, and with the cut-off limit of our samples, we do not probe the regime of late spirals and irregulars.

\begin{table*}
\caption{The $K_s$-band luminosity function parameters of VC04.}
\footnotesize
\begin{center}
\begin{threeparttable}
\begin{tabular}{lccrclc}
\toprule
\toprule
\multicolumn{1}{c}{Cluster}&
\multicolumn{1}{c}{$M^*_{Ks}-5log\,h_{70}$ }&
\multicolumn{1}{c}{$\alpha$} & \multicolumn{1}{c}{$\phi^*$} &
\multicolumn{1}{c}{$r$} &
\multicolumn{1}{c}{Sample Notes}&
\multicolumn{1}{c}{Method} \\
\multicolumn{1}{c}{ }&
\multicolumn{1}{c}{mag}&
\multicolumn{1}{c}{ } & \multicolumn{1}{c}{$h^2_{70}$\,Mpc$^{-2}$} &
\multicolumn{1}{c}{Mpc} &
\multicolumn{1}{c}{}&
\multicolumn{1}{c}{} \\
 \midrule
 VC04 & $-23.72\pm 0.18$ & $-0.75\pm 0.21$ & $16.90\pm 0.10$ & $<1.5$ & $-25\fm5<M_{Ks}<-21\fm5$ & SM\tnote{a} \\ 
 VC04 & $-24.41\pm 0.44$ & $-1.10\pm 0.20$ & $8.84\pm 0.20$ & $<1.5$ & $-25\fm0<M_{Ks}<-21\fm5$ & SM\tnote{a,c} \\  
 VC04 & $-23.70\pm 0.23$ & $-0.50\pm 0.28$ & $13.64\pm 0.12$ & $<1.5$ & $-25\fm5<M_{Ks}<-21\fm5$ & RS\tnote{b} \\ 
 VC04 & $-24.82\pm 0.76$ & $-1.00\pm 0.24$ & $6.11\pm 0.25$ & $<1.5$ & $-25\fm0<M_{Ks}<-21\fm5$ & RS\tnote{b,c} \\ 
 \bottomrule
\end{tabular}
\begin{tablenotes}
\item[a] The spectroscopic membership method.\\
\item[b] The red-sequence method.\\
\item[c] Excluding the first bin.\\
\end{tablenotes}
\end{threeparttable}
\end{center}
\label{tab:lfspropsvc04}
\end{table*}

\begin{figure*}
\centering

        \includegraphics[width=0.7\linewidth]{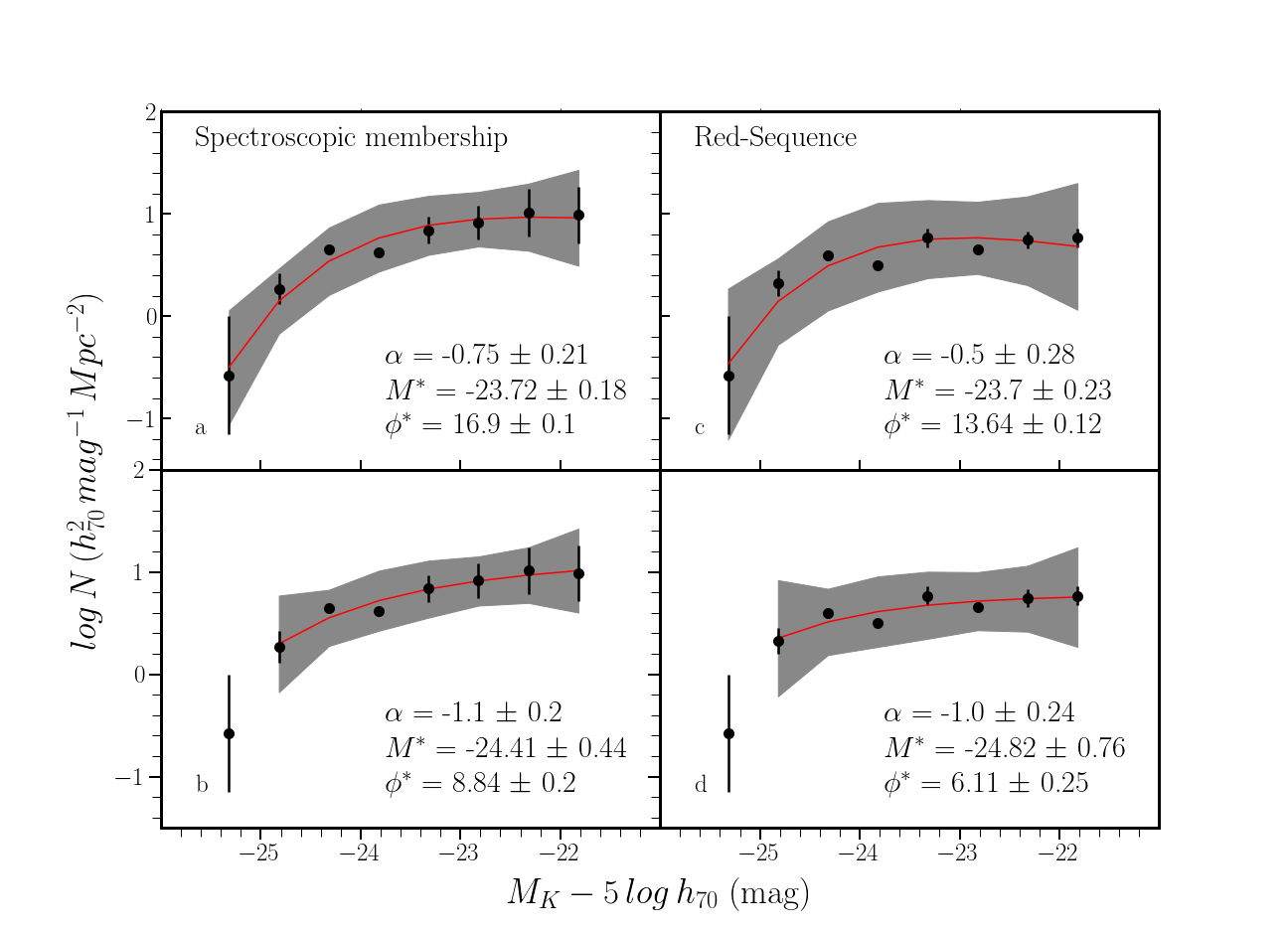}

\caption{\label{fig:lfvc04com}The LFs of the VC04 cluster in the $K_s$-band out to $r_c<1.5~$Mpc. The left panel shows the LFs derived from the spectroscopic method and the right panel derived from the RS method. In both panels the top and bottom plots show the LFs with and without the brightest galaxy respectively. The best fit is shown with the red line and the gray area shows the $3\sigma$ uncertainty of the fit.
}
\end{figure*}

\subsection{Comparison of the two methods}
\label{subsec:methodcom}

Fig.~\ref{fig:lfvc04com} displays the VC04 LFs complete to $r_c<1.5~$Mpc and the extinction-corrected absolute magnitude of $M^o_{Ks}<-21\fm5$, derived from the spectroscopic method in the left panels and the RS method in the right panels. Note that the brightest galaxy of VC04 lies outside of $r<0.5~$Mpc, within a small density enhancement (north-east, see the iso-density contour map of VC04, Fig.~\ref{fig:rsvc04-2}). The redshift distance of this possible subgroup at the outskirts of VC04 is $cz=17873\pm437$~km~s$^{-1}$ based on the available redshift information of galaxies within the small overdensity region, which is very close to the redshift distance of the brightest galaxy ($cz=17827$~km~s$^{-1}$). Therefore, to perform a fair comparison of $r_c<1.5$~Mpc LF of VC04 with the other cluster LFs (where the first bin is often excluded, see Section~\ref{sec:discussion}), the bottom panels of Fig.~\ref{fig:lfvc04com} shows the LF fit with the first bin excluded.  The LF parameters derived for VC04 using both methods are listed in Table~\ref{tab:lfspropsvc04}.

The value of $\phi^*$ is slightly lower for the LF derived using  the RS method compared to the spectroscopic membership method.  
However the VC04 $M^*$ and $\alpha$ extracted from the RS LF are in good agreement (within $1\sigma$) compared to the values derived from the spectroscopic membership method (see Table~\ref{tab:lfspropsvc04}).

The results of the comparison demonstrate that the derivation of the LF based on the RS method is nearly indistinguishable from the spectroscopic one (within $1 \sigma$ uncertainty) and can be reliably used in the absence of high coverage of spectroscopic information.

\subsection{The LFs of other VSCL clusters}

As mentioned before, the redshift data of galaxies in the VSCL clusters (except VC04) were based on spectroscopic data obtained with the 2dF+AAOmega spectrograph on the 3.9m AAT, for 2MASX and optical galaxies. While the VSCL IRSF data are complete to $K_s^o=15\fm5$ ($M_{Ks}^o<-21\fm5$), none of the galaxies fainter than the 2MASX limit ($K_s\ga 13\fm5$, $M_{Ks}\ga -23\fm5$) have redshifts. The fraction of galaxies with spectroscopic data varies in each of the observed VSCL clusters but overall the redshift data coverage is too low to perform the spectroscopic membership method.  The RS method is the only suitable method to derive the LFs of the other four VSCL clusters (VC02, VC05, VC08 and VC11).

\begin{figure*}
\epsscale{0.74}
\plotone{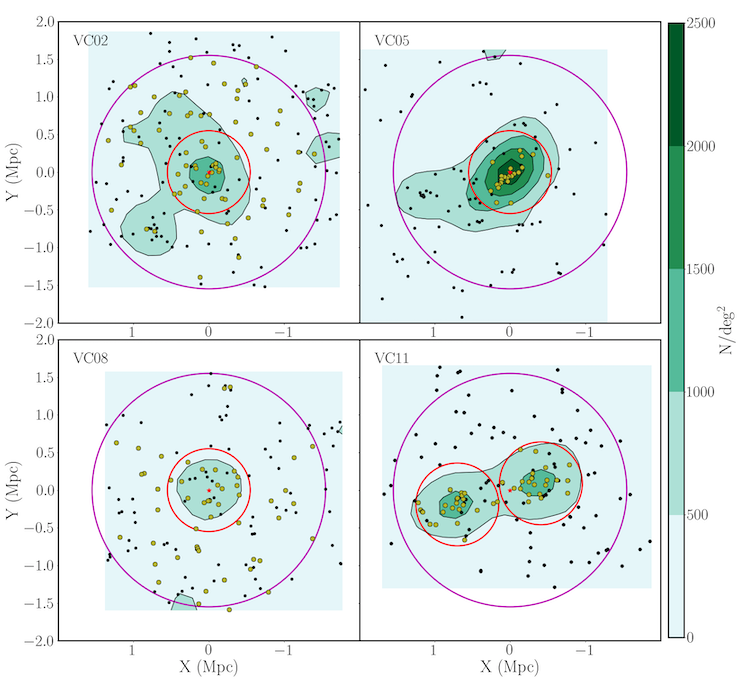}
\caption{The iso-density contour maps of the VC02, VC05, VC08 and VC11 (left clump: VC11a, right clump: VC11b) clusters out to $r_c=1.5\,$Mpc and $M_{Ks}^o<-21\fm5$. The galaxies are shown as black dots and the RS galaxies as yellow dots. The red and magenta circles represent the $r=0.5~$Mpc and $r_c=1.5~$Mpc respectively.}
\label{fig:vsclrs}
\end{figure*}

\begin{figure*}
\epsscale{0.9}
\plotone{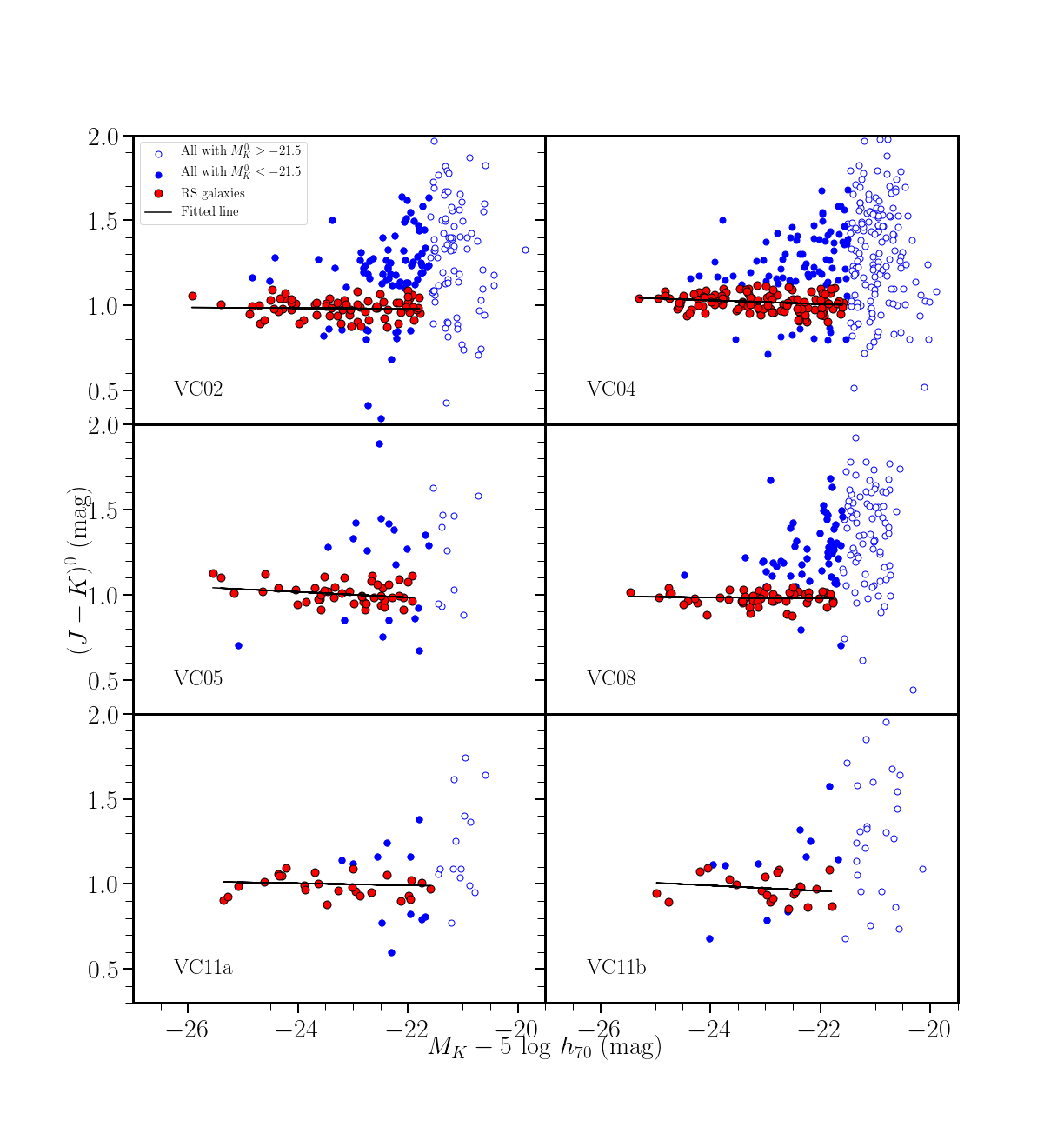}
\caption{\label{fig:vsclrsrelation} The RS of the VSCL clusters performed on galaxies within a clustercentric radii of $r_c<1.5$ for VC02, VC04, VC08 and $r<0.5$ for VC05, VC11a and VC11b. Galaxies with $M_{K_s}^o<-21\fm5$ are shown as filled blue circles and galaxies $M_{K_s}^o>-21\fm5$ as open circles. The RS galaxies are displayed as red circles and the RS fit is shown as black line.
}
\end{figure*}

\begin{figure}
\centering
\includegraphics[width=0.95\linewidth]{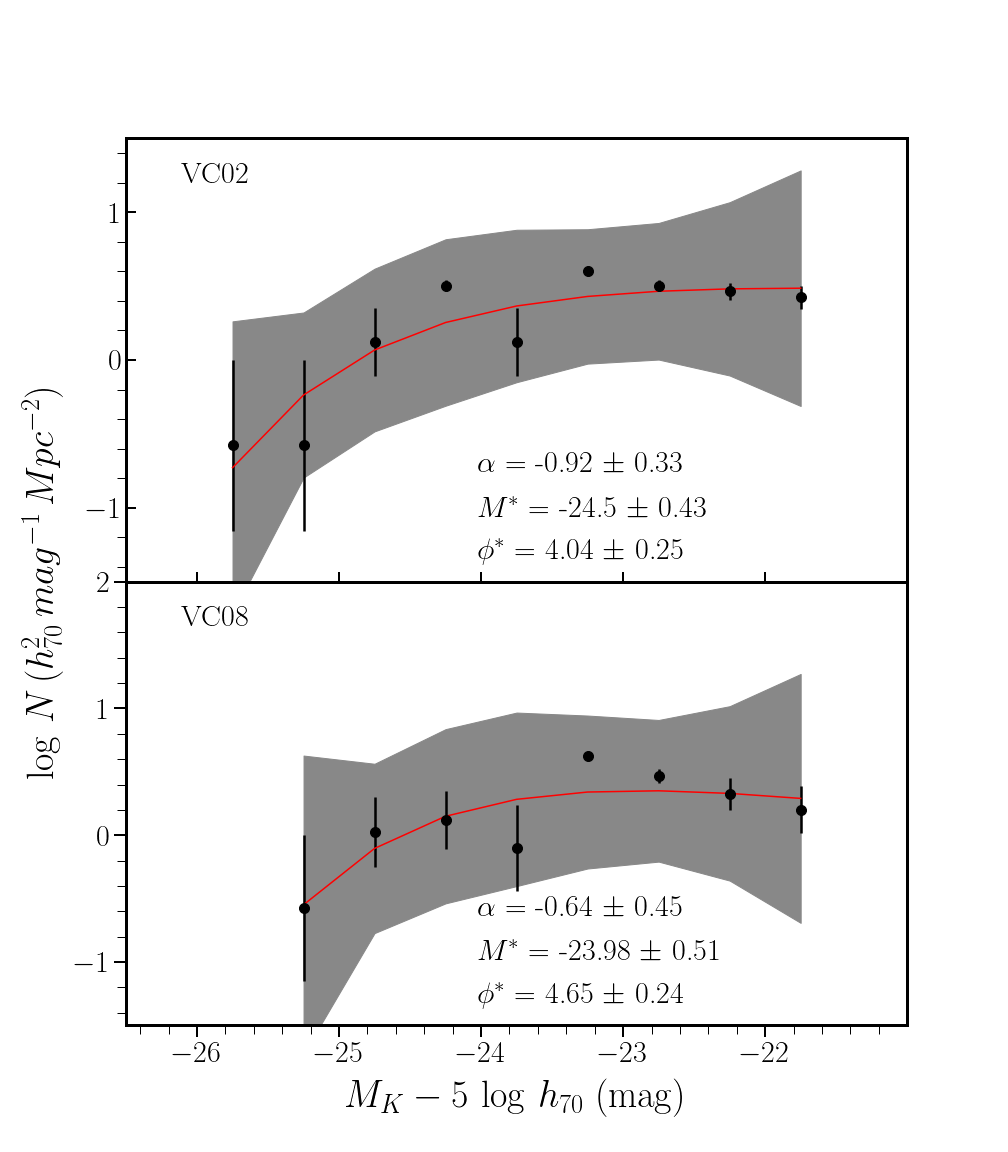}

\caption{\label{fig:vsclrslf}  The $K_s$-band LFs of VC02 and VC08, using the RS method out to $r_c<1.5$\,Mpc. VC02 for $-26\fm0<M_{Ks}^o<-21\fm5$ and VC08 for $-25\fm5<M_{Ks}^o<-21\fm5$. The best fit is shown with the red line and the gray area shows the $3\sigma$ uncertainty of the fit. 
}
\end{figure}

\begin{table}
\caption{\label{tab:rsvscl}RS properties of the VSCL clusters. RS fit relation is defined as $(J-K)^o=(\beta) K^o+ c$. $N_{RS}$ is the number of identified RS galaxies.}
\scriptsize
\hspace*{-1cm}
\begin{tabular}{ccccc}
\toprule
\toprule
\multicolumn{1}{c}{Cluster}&
\multicolumn{1}{c}{$\beta$} &
\multicolumn{1}{c}{$c$}&
\multicolumn{1}{c}{$N_{RS}$ ($1.5~$Mpc)} &
\multicolumn{1}{c}{$N_{RS}$ ($0.5~$Mpc)} \\
 \midrule
 VC02 & $-0.002\pm 0.006$ & $0.936$ & 72 & 23  \\
 VC04 & $-0.011\pm 0.005$ & 0.766 & 118 & 37  \\
 VC05 & $-0.016\pm 0.010$ & $0.632$ & $-$ & 30 \\
 VC08 & $-0.003\pm 0.007$ & $0.912$ & 54 & 20  \\
 VC11a & $-0.006\pm 0.011$ & 0.860 & $-$ & 26 \\
 VC11b & $-0.016\pm 0.019$ & 0.607 & $-$ & 22 \\
 \bottomrule
\end{tabular}
\end{table}
The iso-density contour maps of VC02, VC05, VC08 and VC11 are shown in Fig.~\ref{fig:vsclrs}. Note that VC05 does not have NIR data out to the completeness radius $r_c<1.5~$Mpc because the originally assumed center of the cluster was offset (see Paper~1 for more details) and VC11 contains two subclusters (VC11a, VC11b) and hence is not a centrally concentrated cluster. We therefore, performed the RS method only out to $r<0.5~$Mpc for these clusters (VC05, VC11a and VC11b).  We derived the CMDs of VC02, VC04, VC05, VC08, VC11a and VC11b with their RS fit (see Fig.~\ref{fig:vsclrsrelation}). The resulting RS linear regression and the number of identified RS galaxies in each VSCL cluster is listed in Table~\ref{tab:rsvscl} and plotted as yellow dots in Fig.~\ref{fig:vsclrs}. The number of RS galaxies in VC05, VC11a and VC11b is too low to allow an accurate LF derivation. We therefore, only determined the LFs for the VC02 and VC08 clusters which are fully covered by our photometric survey out to $r_c<1.5~$Mpc using the RS method. Their $K_s$-band LFs are shown in Fig.~\ref{fig:vsclrslf} and the resulting Schechter parameters are listed in Table~\ref{tab:lfvscl}. We summarize the VC02 and VC08 cluster properties and compare their LF parameters with the resulting LF of VC04 in Section~\ref{subsec:env-eff}.

\begin{table*}
\caption{The $K_s$-band luminosity function parameters of the VC02 and VC08 clusters, using the RS method.}
\footnotesize
\begin{center}
\begin{threeparttable}
\begin{tabular}{lccrcl}
\toprule
\toprule
\multicolumn{1}{c}{Cluster}&
\multicolumn{1}{c}{$M^*_{Ks}-5log\,h_{70}$ }&
\multicolumn{1}{c}{$\alpha$} & \multicolumn{1}{c}{$\phi^*$} &
\multicolumn{1}{c}{$r$} &
\multicolumn{1}{c}{Sample Notes} \\
\multicolumn{1}{c}{ }&
\multicolumn{1}{c}{mag}&
\multicolumn{1}{c}{ } & \multicolumn{1}{c}{$h^2_{70}$\,Mpc$^{-2}$} &
\multicolumn{1}{c}{Mpc} &
\multicolumn{1}{c}{} \\
 \midrule
 VC02 & $-24.50\pm 0.43$ & $-0.92\pm 0.33$ & $4.04\pm 0.25$ & $<1.5$ & $-26\fm0<M_{Ks}<-21\fm5$  \\ 
 VC08 & $-23.98\pm 0.51$ & $-0.64\pm 0.45$ & $4.65\pm 0.24$ & $<1.5$& $-25\fm5<M_{Ks}<-21\fm5$  \\ 

 \bottomrule
\end{tabular}
\begin{tablenotes}
\end{tablenotes}
\end{threeparttable}
\end{center}
\label{tab:lfvscl}
\end{table*}

\section{The VC04 LF in the context of other cluster LFs}
\label{sec:discussion}

The Schechter parameters of cluster LFs depend on the wavelength band, magnitude system, location, area and depth of the survey. For example, \citet{andreon2001} found that the same analysis technique yields different LF parameters if applied to different regions of the same cluster. Therefore, useful comparisons between cluster LFs can only be made when the area and depth of the surveys are the same \citep{jerjen1997}. Moreover, because cD galaxies cannot be fitted properly by a Schechter function and not all clusters contain a cD galaxy, it is customary to only compare the Schechter parameters of LFs with the cD galaxies excluded. 

\begin{table*}
\caption{The $K_s$-band luminosity function parameters of Norma, Coma, Virgo and VC04, computed in this work, using the spectroscopic membership method.}
\footnotesize
\begin{center}
\begin{threeparttable}
\begin{tabular}{lccrcl}
\hline
\hline
\multicolumn{1}{c}{Cluster}&
\multicolumn{1}{c}{$M^*_{Ks}-5log\,h_{70}$ }&
\multicolumn{1}{c}{$\alpha$} & \multicolumn{1}{c}{$\phi^*$} &
\multicolumn{1}{c}{$r$} &
\multicolumn{1}{c}{Sample Notes}\\
\multicolumn{1}{c}{ }&
\multicolumn{1}{c}{mag}&
\multicolumn{1}{c}{ } & \multicolumn{1}{c}{$h^2_{70}$\,Mpc$^{-2}$} &
\multicolumn{1}{c}{Mpc} &
\multicolumn{1}{c}{}\\
 \hline
  Coma & $-25.45\pm 0.53$ & $-1.55\pm 0.24$ & $2.38\pm 0.37$ & $<1.5$& $-26\fm5<M_{Ks}<-21\fm5$   \\
 Coma & $-24.33\pm 0.24$ & $-1.14\pm 0.21$ & $9.51\pm 0.17$ & $<1.5$& $-26\fm0<M_{Ks}<-21\fm5$\tnote{a} \\

 \hline
 Norma & $-25.39\pm 0.52$ & $-1.26\pm 0.24$ & $3.50\pm 0.31$ & $<1.5$& $-26\fm5<M_{Ks}<-21\fm5$    \\ 
 Norma & $-24.34\pm 0.21$ & $-0.84\pm 0.17$ & $10.31\pm 0.12$ & $<1.5$& $-26\fm0<M_{Ks}<-21\fm5$\tnote{a}   \\

 \hline
 Virgo & $-25.41\pm 0.36$ & $-1.32\pm 0.11$ & $2.76\pm 0.18$ &  $<1.5$ & $-26\fm0<M_{Ks}<-21\fm5$  \\  
 Virgo & $-24.69\pm 0.28$ & $-1.14\pm 0.12$ & $5.27\pm 0.13$ &  $<1.5$ & $-25\fm5<M_{Ks}<-21\fm5$\tnote{a}  \\ 
 
 \hline
 VC04 & $-23.72\pm 0.18$ & $-0.75\pm 0.21$ & $16.90\pm 0.10$ & $<1.5$ & $-25\fm5<M_{Ks}<-21\fm5$ \\ 
 VC04 & $-24.41\pm 0.44$ & $-1.10\pm 0.20$ & $8.84\pm 0.20$ & $<1.5$ & $-25\fm0<M_{Ks}<-21\fm5$\tnote{a}  \\ \hline 
 
\end{tabular}
\begin{tablenotes}
\item[a] Excluding the first bin which contains the brightest cluster galaxy.
\end{tablenotes}
\end{threeparttable}
\end{center}
\label{tab:complfsprops}
\end{table*}

\subsection{Deriving the Coma, Norma and Virgo LFs}
\label{subsec:discussnorma}
To allow for fair comparisons of the $K_s$-band VSCL LFs with other clusters, we derived the LFs of the Coma, Norma and Virgo clusters using the spectroscopic membership method out to the same completeness magnitude limit ($M_{Ks}^o=-21\fm5$) and clustercentric radius ($r_c=1.5$\,Mpc) as the VSCL survey.
We selected Coma, Norma and Virgo because they are all quite local ($v\la 7000~$km~s$^{-1}$), have NIR ($JHK_s$) photometric data to the approximate same magnitude completeness limit of the more distant VSCL clusters ($\sim 18000~$km~s$^{-1}$) and vary in richness from a young, irregular cluster such as Virgo to a virialized, very massive cluster such as Coma. The list of galaxies used to compute the $K_s$-band LFs of these clusters were taken from the 2MASX catalog \citep{irsa97,jarrett2000A}. 
It should be mentioned that 2MASS is known to miss a considerable fraction of the flux for bright galaxies and to be incomplete for fainter ones \citep{andreon2002,kirby2008}. This incompleteness is  less relevant when working with isophotal magnitudes for normal bulge dominated galaxies. It could affect the dwarf population which we are not probing with our magnitude limit of $-21\fm5$.  We acquired the spectroscopic data of Coma from the SDSS DR16 catalog \citep{ahumada2020}. The redshift information for Norma was obtained  from \citet{woudt2008} and 2MRS  \citep{huchra2012}, and the redshift data for Virgo were obtained from the SDSS DR16 catalog \citep{ahumada2020} and 2MRS \citep{huchra2012}. Our fitted LF parameters of Coma, Norma and Virgo are listed  in Table~\ref{tab:complfsprops}.

Table~\ref{tab:clustercomLF} lists the previously reported Schechter parameters of the Coma and Norma NIR LFs.
\citet{depropris1998} computed the $H$-band LF of the Coma cluster for bright galaxies $M_H<-20\fm5$ using only the cluster members and converted their derived Schechter parameters to the $K$-band applying a color term $(H-K)\sim 0\fm22$. \citet{depropris1998} also  computed the $H$-band LF for faint galaxies in Coma with $-21\fm0<M_H<-19\fm0$ using two methods: the $B-R$ color-magnitude relation and the field correction methods. \citet{andreon2000b} measured the $H$-band LF of galaxies with $M_H<-18\fm0$ using the field correction method. The characteristic $K$-band magnitude for the Coma cluster found by \citet{depropris1998} of $M_K^* = -24\fm02$ agrees well with the value found by \citet{andreon2000b} for the $H$-band ($M_H^* = -23\fm86$), taking into account the color difference $H-K \approx$ 0\fm22 \citep{depropris1998}. No errors for $M^*$ and $\alpha$ are reported in \citet{depropris1998}. Nevertheless the $1\sigma$ error ellipse shown in their Fig.~2 indicates $1\sigma \sim$ 0\fm75 and $1\sigma \sim$ 0.25 for $M^*$ and $\alpha$ respectively. The confidence interval for the LF fit shown in Fig.~3 of \citet{andreon2000b} is even larger.

The Coma $M^*_K$ found in this work is $-24\fm33\pm0.24$ excluding the cD galaxies. This is within $1\sigma$ of what was found in the \citet{depropris1998} and \citet{andreon2000b} studies. The slope $\alpha=-1.14\pm0.21$ lies within $1\sigma$ of \citet{depropris1998}. Their completeness limit, $M_{K}\sim -20\fm5$, goes $1\fm0$ deeper than our work, $M_{K}^o\sim -21\fm5$. For deeper surveys \citep[e.g.][]{andreon2000b} the slope generally is steeper due to the inclusion of the faint dwarf population in their sample.

The Norma NIR LF parameters measured by \citet{skelton2009} are listed in Table~\ref{tab:clustercomLF}. We find excellent agreement with the total LF parameters for the Norma cluster from \citet{skelton2009}. The Norma $M^*_K$ value that we measured in this work ($r_c<1.5$\,Mpc, excluding cD galaxies, $-24\fm34\pm0.21$) is within $1\sigma$ of that of \citet{skelton2009}. However, our slope for Norma ($\alpha=-0.84\pm 0.17$) is shallower than the one found by \citet{skelton2009}; $\alpha=-1.23\pm0.13$ , most likely because their survey extends deeper into the dwarf regime and is sensitive to late spiral and irregular galaxies as well. 

The Schechter parameters have not been published for the Virgo LF in the $K_s$-band to date. \citet{mcdonald2009} plotted the LF of Virgo in the $H$-band but they did not state any parameters for it. Therefore, an external comparison of our Virgo $K_s$-band LF parameters could not be made.

\begin{table*}
\centering
\caption{\label{tab:clustercomLF}Table of previous NIR luminosity functions parameters of clusters. The Norma cluster has two entries in the first and second row of the table which represent the parameters of the Norma LF with and without the first bin (containing the central cD galaxies) respectively.}
\footnotesize
\hspace*{-1.5cm}
\begin{threeparttable}    
\begin{tabular}{lllccl}
\hline
\hline
\multicolumn{1}{c}{Cluster}&
\multicolumn{1}{c}{Reference}&
\multicolumn{1}{c}{Band} & \multicolumn{1}{c}{$M^*$} &
\multicolumn{1}{c}{$\alpha$} &
\multicolumn{1}{c}{Sample Notes}\\
\multicolumn{1}{c}{ }&
\multicolumn{1}{c}{ }&
\multicolumn{1}{c}{ } & \multicolumn{1}{c}{mag} &
\multicolumn{1}{c}{ } &
\multicolumn{1}{c}{ }\\
\hline
Norma & \citet{skelton2009} & $K_s$  & $-25.39\pm0.80$ & $-1.26\pm0.10$ & central 0.8~Mpc$^2$, $-26.0<M_{K_s} < -19.5$\\
Norma & \citet{skelton2009} & $K_s$  & $-25.09\pm1.07$ & $-1.23\pm0.13$ & central 0.8~Mpc$^2$, $-25.5<M_{K_s} < -19.5$\\
\hline
Coma & \citet{depropris1998} & $H$ & -23.82 & -0.93 & central 0.53~Mpc$^2$,  $M_H < -20.5$\\
Coma & \citet{depropris1998} & $K$ & -24.02 & -0.98 & central 0.53~Mpc$^2$, $M_H < -20.5$\\
Coma & \citet{andreon2000b} & $H$  & -23.86 & -1.3 & off-center 0.30~Mpc$^2$, $M_H < -18.0$\\
\hline
AC118 & \citet{andreon2001} & $K_s$ & -25.26 & -1.2 & 1.36\,Mpc$^2$, $M_{K}\la-21.5$\\
AC118 &  \citet{andreon2001} & $K_s$ & -23.96 & -0.5 & 0.26\,Mpc$^2$ Main clump\\
AC118 &  \citet{andreon2001} & $K_s$ & -23.56 & -0.9 & 0.26\,Mpc$^2$ NW clump\\
\hline
93 clusters\tnote{a} & \citet{lin2004} & $K_s$ & $-24.34\la M^* \la -24.02$ & $-1.1\la \alpha \la -0.84$ & Composite LF, $M_{K}\la-21.0$ \\
\hline
5 clusters\tnote{b} & \citet{merluzzi2010} & $K_s$ &$-24.96 \pm 0.10$ & $-1.42 \pm 0.03$ & Composite LF, $M_{K}\la-18.5$\\	
\hline
24 clusters\tnote{c} & \citet{depropris2017} & $K$ &$-24.78 \pm 0.14$ & $-1.41 \pm 0.10$ & Composite LF\tnote{d} , $M_{K}\la-22.0$ \\
24 clusters & \citet{depropris2017} & $K$ &$-24.41 \pm 0.15$ & $-1.00 \pm 0.12$ & Composite LF\tnote{e}\\
\hline

\end{tabular}

\footnotesize{The $M^*$s have been adjusted for H$_0$ = 70~km~s$^{-1}$~Mpc$^{-1}$. The difference between the magnitude systems have not been taken into account, however characteristic magnitudes in $H$ and $K$-bands can be converted using the color: $H - K \sim 0\fm22$ \citep{depropris1998}.}
\begin{tablenotes}
\item[a] $0.01\la z\la 0.09$.\\
\item[b] Clusters in the Shapley supercluster at $z=0.048$.\\
\item[c] Clusters are at mean redshift of $z=0.075.$\\
\item[d] Using the spectroscopic information.\\
\item[e] The RS LFs.
\end{tablenotes}
\end{threeparttable}
\end{table*}

\begin{figure}
\centering

\includegraphics[width=1.05\linewidth]{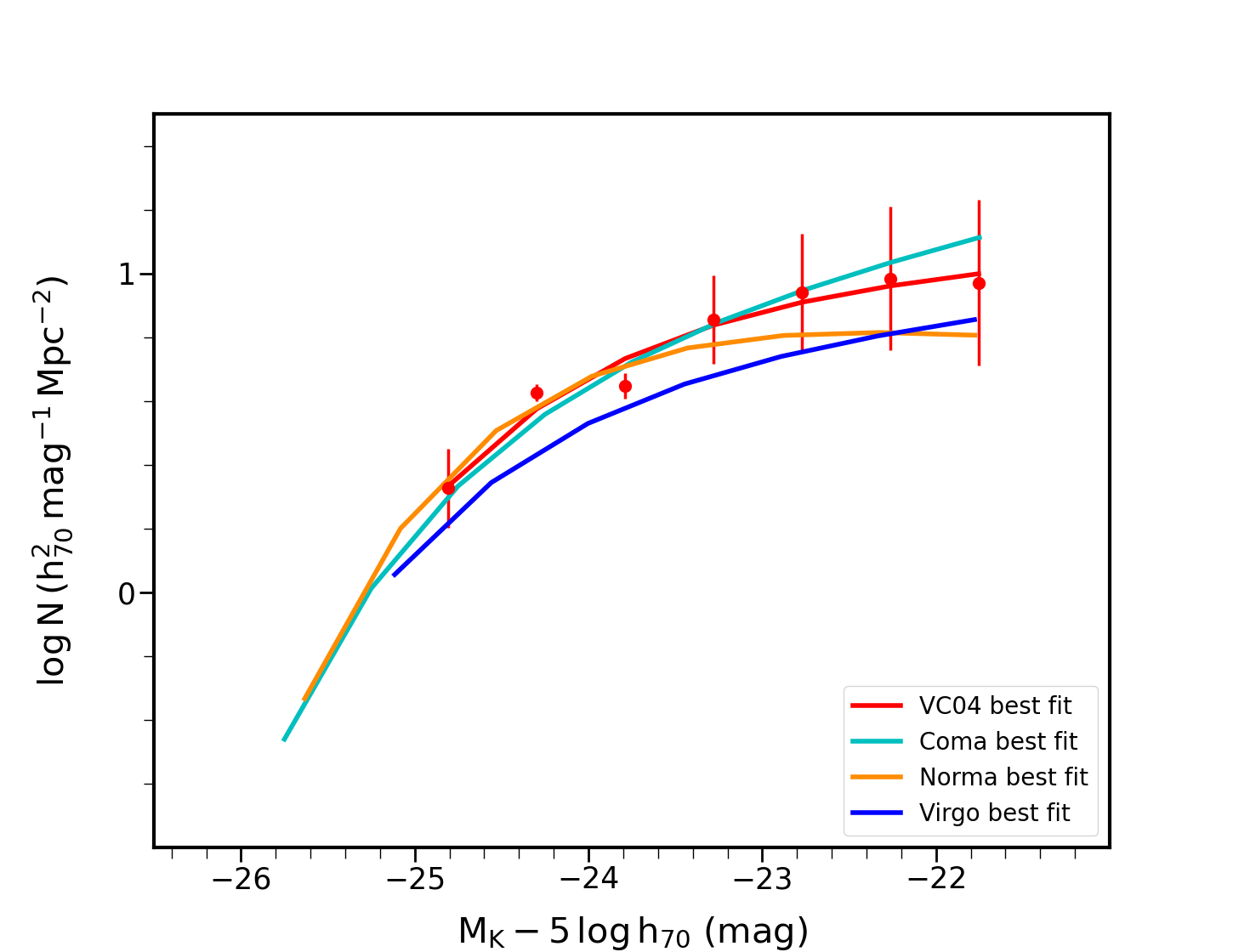}
\includegraphics[width=1.05\linewidth]{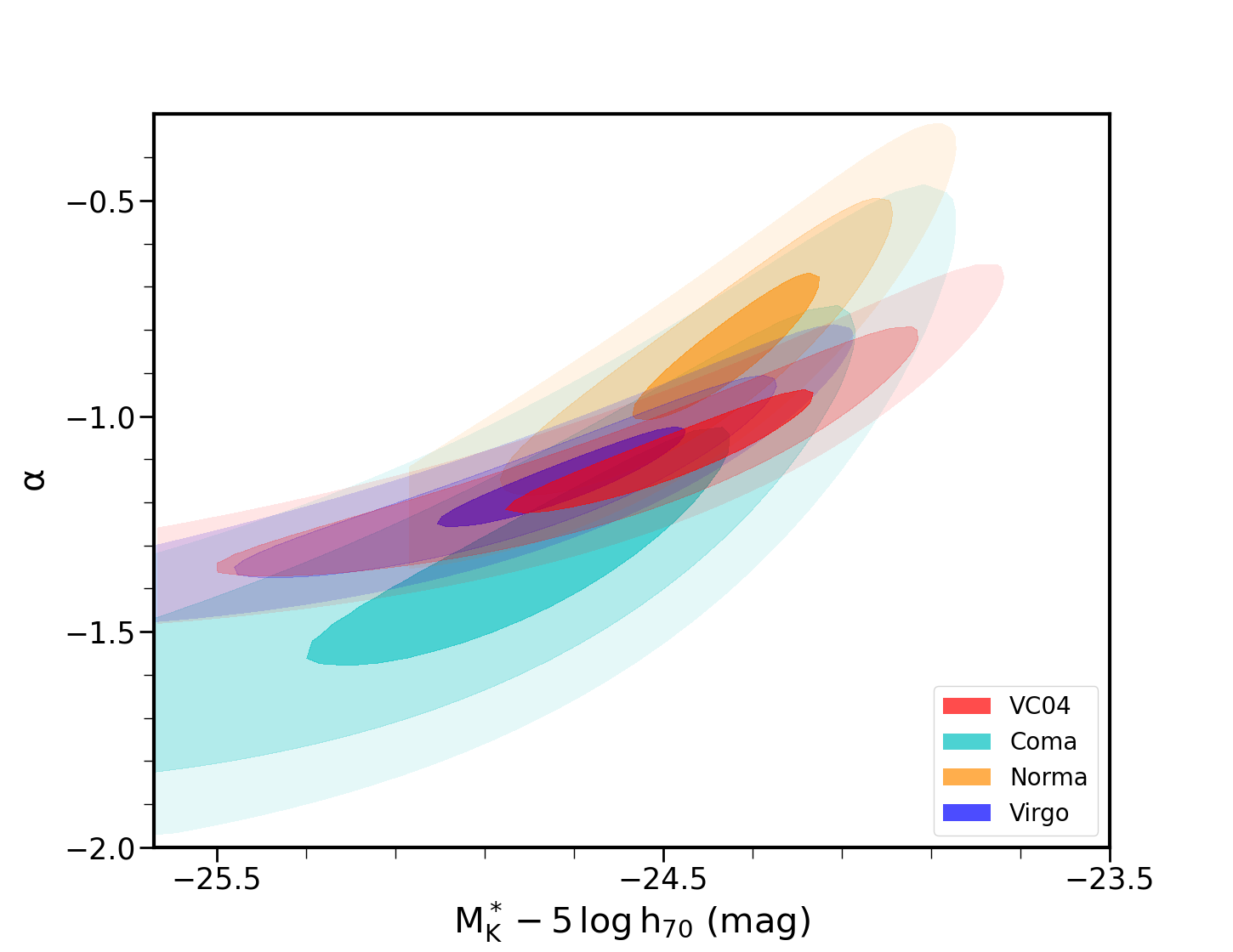}
\caption{\label{fig:all-lf} The $K_s$-band LFs and best fit Schechter functions of the VC04 (red), Coma (cyan), Norma (orange) and Virgo (blue) clusters out to $r_c<1.5$\,Mpc and $M_{Ks}^o<-21\fm5$. The associated 1, 2 and 3$\sigma$ error ellipses for the best-fitting LF parameters are shown in the bottom panel. Error contours have the same color as their parent LFs. 
}
\end{figure}

\begin{table*}
\caption{The $K_s$-band Luminosity Function parameters of VC02 and VC08 for a fixed $M^*=-24\fm41$, using the RS method.}
\footnotesize
\begin{center}
\begin{tabular}{lccrcl}
\toprule
\toprule
\multicolumn{1}{c}{Cluster}&
\multicolumn{1}{c}{$M^*_{Ks}-5log\,h_{70}$ }&
\multicolumn{1}{c}{$\alpha$} &
\multicolumn{1}{c}{$\phi^*$} &
\multicolumn{1}{c}{$r$} &
\multicolumn{1}{c}{Sample Notes} \\
\multicolumn{1}{c}{ }&
\multicolumn{1}{c}{mag}&
\multicolumn{1}{c}{ } &
\multicolumn{1}{c}{$h^2_{70}$\,Mpc$^{-2}$} &
\multicolumn{1}{c}{Mpc} &
\multicolumn{1}{c}{} \\
 \midrule
 VC02  & $-24\fm41$ & $-0.78\pm0.16$ &$5.17\pm 0.10$ & $<1.5$ & $-25\fm0<M_{Ks}<-21\fm5$  \\
 VC08  & $-24\fm41$ & $-0.83\pm0.22$ & $3.44\pm 0.13$ & $<1.5$ & $-25\fm0<M_{Ks}<-21\fm5$  \\
 \bottomrule
\end{tabular}

\end{center}
\label{tab:lfvsclfixed}
\end{table*}

\subsection{The VC04 cluster LF compared to Coma, Norma and Virgo}

We compare the VC04 LF parameters with the parameters we measured for Coma, Norma and Virgo. All comply with the same magnitude and radius completeness limit. The LFs of VC04, Coma, Norma and Virgo (excluding their first bin) are shown in Fig.~\ref{fig:all-lf} together with the corresponding error ellipses. The $M^*_K$ of the VC04 LF ($-24\fm41\pm0.44$, excluding the first bin) lies within the range of the Norma, Coma and Virgo clusters.  The slope of the VC04 LF ($-1.10\pm0.20$) is in good agreement with those of the Coma and Virgo cluster, and only slightly steeper than the slope of the Norma LF. It is clear in Fig.~\ref{fig:all-lf} that the $M^*_K$ and $\alpha$ values of the clusters are strongly correlated.

The Coma cluster contains substructure \citep{biviano1996,healy2021c} and could therefore, have more late-type galaxies in its outer regions while Virgo is a young and intermediate cluster with many late-type galaxies. \citet{woudt2008} shows that there are two subgroups in the Norma cluster that are populated by spiral/irregular galaxies. Note that the more prominent subgroup of Norma is located roughly at the border of the completeness radius of this survey ($r_c<1.5~$Mpc) and it could be an infalling group. Based on the derived slope of the NIR LFs in our work, it seems that VC04 contains more late-type galaxies than Norma for $M_{Ks}^o<-21\fm5$ resulting in the steeper slope of its LF.  
A comparison between the volume density $\phi^*$ of the clusters indicates that the number density of VC04 agrees well with those of the Coma and Norma clusters and is higher than the Virgo LF $\phi^*$. This seems to indicate that VC04 is a massive cluster, but not yet fully relaxed.

It should be noted that the $M^*_K$ and faint end slope of the VC02 LF ($-24\fm50\pm0.43$ and $-0.92\pm0.33$ respectively) are within $1\sigma$ of the derived values for the Coma, Norma and Virgo clusters. The $M^*_K$ of the VC08 LF ($-23\fm98\pm0.51$) is in agreement with those of the Coma and Norma clusters and only slightly brighter than the Virgo $M^*_K$ but VC08 has similar slope ($-0.64\pm0.45$) compared to the Coma, Norma and Virgo LFs (see Tables~\ref{tab:lfvscl} and \ref{tab:complfsprops}).

\subsection{Comparison with the published NIR LFs}

Table~\ref{tab:clustercomLF} lists some of the other previous NIR cluster and composite LFs parameters.  \citet{lin2004,merluzzi2010} and \citet{depropris2017} computed the composite LFs of clusters at nearby redshifts as listed in Table~\ref{tab:clustercomLF}.
The $M^*_{K}$ of VC04 LF ($M^*_{K}=-24\fm41\pm 0.44$) is in good agreement with that of the previous $K$-band cluster LF studies in the literature (considering their respective uncertainties), for samples of similar depth. 
\citet{lin2004} computed a composite LF with $-24\fm34\la M^*_K \la -24\fm02$ for 93 clusters at $0.01\la z \la 0.09$ out to $M_K\la -21\fm0$.
For 24 clusters at $z=0.075$,  \citet{depropris2017} derived $M^*_K=-24\fm78\pm 0.14$  using the spectroscopic information, and $M^*_K=-24\fm41\pm 0.15$ using the RS members.
\citet{merluzzi2010} measured $M^*_K=-24\fm96\pm 0.10$ for composite LF of 5 clusters in the Shapley supercluster complete to $M_K\la -18\fm5$. Their $M^*_K$ is only slightly brighter than the $M^*_K$ of the VC04 LF.

The faint end slope of the VC04 LF ($\alpha=-1.10\pm 0.20$) is in agreement with that of \citet{lin2004} ($-1.1\la \alpha \la -0.84$) and the RS composite LF of \citet{depropris2017} ($-1.0\pm 0.12$), but is shallower than the version derived by 
\citet{depropris2017} when they only used the spectroscopic information ($-1.41\pm 0.10$), despite their similar magnitude range to our work ($M_{K}\la-22\fm0$). The faint end slope determined by \citet{depropris2017} for their full sample of red and blue galaxies in 24 clusters  ($-1.41\pm 0.10$) is also steeper than the value they derived in their previous analysis of a subset of these clusters \citep[$-0.98$,][]{depropris2009}.  
\citet{merluzzi2010}'s survey is deeper than this work and therefore, the upturn in their composite LFs, caused by the power law curve of the late-type and dwarf galaxies, is likely the reason for their steeper faint end slope ($-1.42\pm 0.03$).

AC118 is a cluster at $z=0.3$ and comparison with LFs of nearby clusters could provide information on the evolution of galaxies in clusters. \citet{andreon2001} found $\alpha=-1.2$ and $M^*_K=-25\fm26$ for the $K$-band LF of AC118 up to $M_{K}\la -21\fm5$ (same as this work). The slope of the AC118 LF is within $1\sigma$ uncertainty of the VC04 faint end slope, however its $M^*_K$ is brighter than that of the VC04 LF. This might be due to evolution of galaxies in clusters from intermediate redshifts to the present time. 
\citet{andreon2001} showed that in their sample the number of bright early-type galaxies in the clump is higher (due to its denser environment) compared to the number of dwarfs (which is higher in the outer regions).

\subsection{Environmental effects}
\label{subsec:env-eff}

\citet{merluzzi2010} studied the LF of clusters in the Shapley supercluster in three regions with different densities of galaxies and found that the faint-end slope increases significantly as one moves from high-density to low-density regions and the upturn in the LF for $M_K>-21\fm0$ becomes more apparent in low-density environments. \citet{depropris2017} found that $M_{K}^*$ is largely unaffected by the environment but that the faint-end slope, $\alpha$, increases towards lower mass clusters ($\sigma < 800$~km~s$^{-1}$) and clusters with Bautz-Morgan type below II. Their RS LF appears, however, to be unaffected by the environment.  A steeper faint-end slope in low-density regions has been observed in optical bands as well \citep[e.g. ][]{boue2008,McNaught2014,lee2016}. 

In Paper~1 we showed that VC04 is the richest cluster of the five VSCL clusters with velocity dispersion, $\sigma_v=515~$km~s$^{-1}$. VC02 at $cz=18955~$km~s$^{-1}$ with $\sigma_v=670~$km~s$^{-1}$ is a relatively rich cluster, and VC08 at $cz=17298~$km~s$^{-1}$ with $\sigma_v=355~$km~s$^{-1}$ is a poor cluster. 
The LF parameters of VC02 ($M^*_K=-24\fm50\pm0.43$ and $\alpha= -0.92\pm0.33$) and VC08 ($M^*_K=-23\fm98\pm0.51$ and $\alpha= -0.64\pm0.45$) are in agreement with VC04 LF parameters ($M^*_K=-24\fm41\pm0.44$ and $\alpha= -1.10\pm0.20$) considering their respective uncertainties. The morphological mix of galaxies in the VSCL clusters is similar out to the completeness radius $r_c<1.5~$Mpc and magnitude limit ($M_{Ks}<-21\fm5$) of this work. To allow for a fair comparison we also derived the $\alpha$ and $\phi^*$ of VC02 and VC08 for a fixed $M^*_K=-24\fm41$ which is the $M^*_K$ of the VC04 cluster for $r_c<1.5~$Mpc, and magnitude range of $-25\fm0<M_{Ks}<-21\fm5$. The results are listed in Table~\ref{tab:lfvsclfixed} and show that the resulting $\alpha$ values for VC02 and VC08 are still in good agreement with what is found for VC04 considering their uncertainties.  Note that VC02, VC04 and VC08 have velocity dispersions of $\sigma_v < 800$~km~s$^{-1}$. However, the VC04 LF parameters are consistent with those of the rich Coma and Norma clusters with $\sigma_v > 800$~km~s$^{-1}$ (see Section~\ref{subsec:discussnorma}).


\section{Summary and results}
\label{sec:summary}
We investigated the properties of galaxy clusters in the VSCL by measuring their LFs. The NIR dataset is complete to $r_c<1.5$~Mpc for $M_{Ks}^o<-21\fm5$. Our main focus is the VC04 cluster, the richest of the five clusters in the VSCL with deep NIR data (Paper~1). For this cluster we acquired new spectroscopic data using SALT that allows the determination of redshifts to  considerably fainter magnitudes than other VSCL clusters with existing spectroscopic data that did not probe fainter than $M_{Ks}^o<-24\fm0$. 

We use the VC04 dataset to demonstrate that the RS method is as reliable a method to determine cluster membership and derive cluster LFs as the spectroscopic membership method. Based on this result we applied the RS method to derive the LFs of the other VSCL clusters which do not have a high spectroscopic data coverage. 
For VC04, we obtained $M^*_K=-24\fm41\pm 0.44$, $\alpha=-1.10\pm 0.20$, $\phi^*=8.84\pm 0.20$ using the spectroscopic membership method.
The VC04 LF parameters were measured for a sample complete to $M_{Ks}^o<-21\fm5$, $\sim 2\fm5$ deeper than $M^*_K$. It therefore, is mainly sensitive to ellipticals and bulgy spirals, with Gaussian contributions to the shape of the LF. There is no upturn at the faint end of the derived LF because the NIR sample does not reach the dwarf regime. We also derived the $K$-band LFs of the Coma, Norma and Virgo clusters out to the same radius and magnitude completeness limit of the VSCL clusters to allow for fair comparisons of their LFs. The LF comparisons  showed that $M^*_K$ and $\alpha$ of the VC04 LF agree well (within errors) with those of other local clusters (Coma, Norma and Virgo) and the other two VSCL clusters (VC02 and VC08) with different masses, structures and galaxy densities. However, the volume density $\phi^*$ of VC04 agrees well with those of the Coma and Norma clusters and is higher than the Virgo, VC02 and VC08 LF $\phi^*$.
The LF parameters of VC04 were also compared to those of the AC118 cluster at $z=0.3$ \citep{andreon2001}. The VC04 LF has a fainter $M^*_K$ compared to that of AC118 LF which might be due to evolution of galaxies in clusters from intermediate redshifts to the present time. The AC118 sample which is complete to the same magnitude limit as our NIR sample, contains a large number of dwarfs in the outer regions of the cluster. 

The new spectroscopic information of VC04 (listed in Table.~\ref{tab:redshift-catalog}) and its LF confirms VC04 to be a massive cluster which is not yet fully relaxed.
The good agreement between the VC04 LF parameters and the other local as well as VSCL clusters, probing out to large clustercentric distances ($\sim 7.5$~Mpc$^2$), suggest the $K$-band LF to be unaffected by the environment at $M_K^o<M^*_K+2\fm5$.

We found that the NIR RS is a valuable tool in identifying cluster members and the determination of NIR LFs of clusters with limited spectroscopy. RS galaxies can therefore be reliably  used for the analysis of the richness of clusters in the VSCL and better understanding of its properties. Our analysis of five of clusters in the VSCL suggests it to be a rich supercluster with a large number of massive clusters. However, more analyzes need to be performed on the rest of the VSCL clusters to prove the hypothesis that the VSCL is indeed a massive and extended supercluster.

\section{Acknowledgments}
\begin{acknowledgments}
We would like to acknowledge the support from the Science Faculty at University of Cape Town (UCT), the South African Research Chairs Initiative of the Department of Science and Technology and the South African National Research Foundation. Some of the observations reported in this paper were obtained with the Southern African Large Telescope (SALT). This paper uses observations made at the South African Astronomical Observatory (SAAO). This publication makes use of data products from the Two Micron All Sky Survey, which is a joint project of the University of Massachusetts and the Infrared Processing and Analysis Center/ California Institute of Technology, funded by the National Aeronautics and Space Administration and the National Science Foundation. We would like to thank Prof. Matt Hilton for his great help with the SALT MOS pipeline.
\end{acknowledgments}

%

\vspace{5mm}
\facilities{IRSF, SALT}









\bibliography{LFVSCL}{}

\begin{thebibliography}{}
\expandafter\ifx\csname natexlab\endcsname\relax\def\natexlab#1{#1}\fi
\providecommand{\url}[1]{\href{#1}{#1}}
\providecommand{\dodoi}[1]{doi:~\href{http://doi.org/#1}{\nolinkurl{#1}}}
\providecommand{\doeprint}[1]{\href{http://ascl.net/#1}{\nolinkurl{http://ascl.net/#1}}}
\providecommand{\doarXiv}[1]{\href{https://arxiv.org/abs/#1}{\nolinkurl{https://arxiv.org/abs/#1}}}

\bibitem[{{Abell}(1975)}]{abell1975}
{Abell}, G.~O. 1975, in Galaxies and the Universe, ed. A.~{Sandage}, M.~{Sandage}, \& J.~{Kristian}, 601

\bibitem[{{Ahumada} {et~al.}(2020){Ahumada}, {Prieto}, {Almeida}, {Anders}, {Anderson}, {Andrews}, {Anguiano}, {Arcodia}, {Armengaud}, {Aubert}, {Avila}, {Avila-Reese}, {Badenes}, {Balland}, {Barger}, {Barrera-Ballesteros}, {Basu}, {Bautista}, {Beaton}, {Beers}, {Benavides}, {Bender}, {Bernardi}, {Bershady}, {Beutler}, {Bidin}, {Bird}, {Bizyaev}, {Blanc}, {Blanton}, {Boquien}, {Borissova}, {Bovy}, {Brandt}, {Brinkmann}, {Brownstein}, {Bundy}, {Bureau}, {Burgasser}, {Burtin}, {Cano-D{\'\i}az}, {Capasso}, {Cappellari}, {Carrera}, {Chabanier}, {Chaplin}, {Chapman}, {Cherinka}, {Chiappini}, {Doohyun Choi}, {Chojnowski}, {Chung}, {Clerc}, {Coffey}, {Comerford}, {Comparat}, {da Costa}, {Cousinou}, {Covey}, {Crane}, {Cunha}, {Ilha}, {Dai}, {Damsted}, {Darling}, {Davidson}, {Davies}, {Dawson}, {De}, {de la Macorra}, {De Lee}, {Queiroz}, {Deconto Machado}, {de la Torre}, {Dell'Agli}, {du Mas des Bourboux}, {Diamond-Stanic}, {Dillon}, {Donor}, {Drory}, {Duckworth}, {Dwelly}, {Ebelke}, {Eftekharzadeh}, {Davis
  Eigenbrot}, {Elsworth}, {Eracleous}, {Erfanianfar}, {Escoffier}, {Fan}, {Farr}, {Fern{\'a}ndez-Trincado}, {Feuillet}, {Finoguenov}, {Fofie}, {Fraser-McKelvie}, {Frinchaboy}, {Fromenteau}, {Fu}, {Galbany}, {Garcia}, {Garc{\'\i}a-Hern{\'a}ndez}, {Oehmichen}, {Ge}, {Maia}, {Geisler}, {Gelfand}, {Goddy}, {Gonzalez-Perez}, {Grabowski}, {Green}, {Grier}, {Guo}, {Guy}, {Harding}, {Hasselquist}, {Hawken}, {Hayes}, {Hearty}, {Hekker}, {Hogg}, {Holtzman}, {Horta}, {Hou}, {Hsieh}, {Huber}, {Hunt}, {Chitham}, {Imig}, {Jaber}, {Angel}, {Johnson}, {Jones}, {J{\"o}nsson}, {Jullo}, {Kim}, {Kinemuchi}, {Kirkpatrick}, {Kite}, {Klaene}, {Kneib}, {Kollmeier}, {Kong}, {Kounkel}, {Krishnarao}, {Lacerna}, {Lan}, {Lane}, {Law}, {Le Goff}, {Leung}, {Lewis}, {Li}, {Lian}, {Lin}, {Long}, {Longa-Pe{\~n}a}, {Lundgren}, {Lyke}, {Ted Mackereth}, {MacLeod}, {Majewski}, {Manchado}, {Maraston}, {Martini}, {Masseron}, {Masters}, {Mathur}, {McDermid}, {Merloni}, {Merrifield}, {M{\'e}sz{\'a}ros}, {Miglio}, {Minniti}, {Minsley}, {Miyaji},
  {Mohammad}, {Mosser}, {Mueller}, {Muna}, {Mu{\~n}oz-Guti{\'e}rrez}, {Myers}, {Nadathur}, {Nair}, {Nandra}, {do Nascimento}, {Nevin}, {Newman}, {Nidever}, {Nitschelm}, {Noterdaeme}, {O'Connell}, {Olmstead}, {Oravetz}, {Oravetz}, {Osorio}, {Pace}, {Padilla}, {Palanque-Delabrouille}, {Palicio}, {Pan}, {Pan}, {Parker}, {Paviot}, {Peirani}, {Ram{\'r}ez}, {Penny}, {Percival}, {Perez-Fournon}, {P{\'e}rez-R{\`a}fols}, {Petitjean}, {Pieri}, {Pinsonneault}, {Poovelil}, {Povick}, {Prakash}, {Price-Whelan}, {Raddick}, {Raichoor}, {Ray}, {Rembold}, {Rezaie}, {Riffel}, {Riffel}, {Rix}, {Robin}, {Roman-Lopes}, {Rom{\'a}n-Z{\'u}{\~n}iga}, {Rose}, {Ross}, {Rossi}, {Rowlands}, {Rubin}, {Salvato}, {S{\'a}nchez}, {S{\'a}nchez-Menguiano}, {S{\'a}nchez-Gallego}, {Sayres}, {Schaefer}, {Schiavon}, {Schimoia}, {Schlafly}, {Schlegel}, {Schneider}, {Schultheis}, {Schwope}, {Seo}, {Serenelli}, {Shafieloo}, {Shamsi}, {Shao}, {Shen}, {Shetrone}, {Shirley}, {Aguirre}, {Simon}, {Skrutskie}, {Slosar}, {Smethurst}, {Sobeck}, {Sodi},
  {Souto}, {Stark}, {Stassun}, {Steinmetz}, {Stello}, {Stermer}, {Storchi-Bergmann}, {Streblyanska}, {Stringfellow}, {Stutz}, {Su{\'a}rez}, {Sun}, {Taghizadeh-Popp}, {Talbot}, {Tayar}, {Thakar}, {Theriault}, {Thomas}, {Thomas}, {Tinker}, {Tojeiro}, {Toledo}, {Tremonti}, {Troup}, {Tuttle}, {Unda-Sanzana}, {Valentini}, {Vargas-Gonz{\'a}lez}, {Vargas-Maga{\~n}a}, {V{\'a}zquez-Mata}, {Vivek}, {Wake}, {Wang}, {Weaver}, {Weijmans}, {Wild}, {Wilson}, {Wilson}, {Wolthuis}, {Wood-Vasey}, {Yan}, {Yang}, {Y{\`e}che}, {Zamora}, {Zarrouk}, {Zasowski}, {Zhang}, {Zhao}, {Zhao}, {Zheng}, {Zheng}, {Zhu}, \& {Zou}}]{ahumada2020}
{Ahumada}, R., {Prieto}, C.~A., {Almeida}, A., {et~al.} 2020, \apjs, 249, 3, \dodoi{10.3847/1538-4365/ab929e}

\bibitem[{{Andreon}(1998)}]{andreon1998}
{Andreon}, S. 1998, \aap, 336, 98.
\newblock \doarXiv{astro-ph/9804028}

\bibitem[{{Andreon}(2001)}]{andreon2001}
---. 2001, \apj, 547, 623, \dodoi{10.1086/318381}

\bibitem[{{Andreon}(2002)}]{andreon2002}
---. 2002, \aap, 382, 495, \dodoi{10.1051/0004-6361:20011636}

\bibitem[{{Andreon} {et~al.}(1997){Andreon}, {Davoust}, \& {Heim}}]{andreon1997}
{Andreon}, S., {Davoust}, E., \& {Heim}, T. 1997, \aap, 323, 337.
\newblock \doarXiv{astro-ph/9612027}

\bibitem[{{Andreon} \& {Pell{\'o}}(2000)}]{andreon2000b}
{Andreon}, S., \& {Pell{\'o}}, R. 2000, \aap, 353, 479.
\newblock \doarXiv{astro-ph/9911329}

\bibitem[{{Andreon} {et~al.}(2005){Andreon}, {Punzi}, \& {Grado}}]{andreon2005}
{Andreon}, S., {Punzi}, G., \& {Grado}, A. 2005, \mnras, 360, 727, \dodoi{10.1111/j.1365-2966.2005.09063.x}

\bibitem[{{Bell} \& {de Jong}(2001)}]{bell2001}
{Bell}, E.~F., \& {de Jong}, R.~S. 2001, \apj, 550, 212, \dodoi{10.1086/319728}

\bibitem[{{Bell} {et~al.}(2003){Bell}, {McIntosh}, {Katz}, \& {Weinberg}}]{bell2003}
{Bell}, E.~F., {McIntosh}, D.~H., {Katz}, N., \& {Weinberg}, M.~D. 2003, \apjs, 149, 289, \dodoi{10.1086/378847}

\bibitem[{{Benson} {et~al.}(2003){Benson}, {Bower}, {Frenk}, {Lacey}, {Baugh}, \& {Cole}}]{benson2003}
{Benson}, A.~J., {Bower}, R.~G., {Frenk}, C.~S., {et~al.} 2003, \apj, 599, 38, \dodoi{10.1086/379160}

\bibitem[{{Binggeli} {et~al.}(1988){Binggeli}, {Sandage}, \& {Tammann}}]{binggeli1988}
{Binggeli}, B., {Sandage}, A., \& {Tammann}, G.~A. 1988, \araa, 26, 509, \dodoi{10.1146/annurev.aa.26.090188.002453}

\bibitem[{{Biviano} {et~al.}(1995){Biviano}, {Durret}, {Gerbal}, {Le Fevre}, {Lobo}, {Mazure}, \& {Slezak}}]{biviano1995}
{Biviano}, A., {Durret}, F., {Gerbal}, D., {et~al.} 1995, \aap, 297, 610.
\newblock \doarXiv{astro-ph/9411025}

\bibitem[{{Biviano} {et~al.}(1996){Biviano}, {Durret}, {Gerbal}, {Le Fevre}, {Lobo}, {Mazure}, \& {Slezak}}]{biviano1996}
---. 1996, \aap, 311, 95.
\newblock \doarXiv{astro-ph/9512111}

\bibitem[{{Blakeslee} {et~al.}(2003){Blakeslee}, {Franx}, {Postman}, {Rosati}, {Holden}, {Illingworth}, {Ford}, {Cross}, {Gronwall}, {Ben{\'\i}tez}, {Bouwens}, {Broadhurst}, {Clampin}, {Demarco}, {Golimowski}, {Hartig}, {Infante}, {Martel}, {Miley}, {Menanteau}, {Meurer}, {Sirianni}, \& {White}}]{blakeslee2003}
{Blakeslee}, J.~P., {Franx}, M., {Postman}, M., {et~al.} 2003, \apjl, 596, L143, \dodoi{10.1086/379234}

\bibitem[{{Bou{\'e}} {et~al.}(2008){Bou{\'e}}, {Adami}, {Durret}, {Mamon}, \& {Cayatte}}]{boue2008}
{Bou{\'e}}, G., {Adami}, C., {Durret}, F., {Mamon}, G.~A., \& {Cayatte}, V. 2008, \aap, 479, 335, \dodoi{10.1051/0004-6361:20077723}

\bibitem[{{Bower} {et~al.}(1992){Bower}, {Lucey}, \& {Ellis}}]{bower1992}
{Bower}, R.~G., {Lucey}, J.~R., \& {Ellis}, R.~S. 1992, \mnras, 254, 601, \dodoi{10.1093/mnras/254.4.601}

\bibitem[{{Bromley} {et~al.}(1998){Bromley}, {Press}, {Lin}, \& {Kirshner}}]{bromley1998}
{Bromley}, B.~C., {Press}, W.~H., {Lin}, H., \& {Kirshner}, R.~P. 1998, \apj, 505, 25, \dodoi{10.1086/306144}

\bibitem[{{Buckley} {et~al.}(2006){Buckley}, {Swart}, \& {Meiring}}]{buckley2006}
{Buckley}, D.~A.~H., {Swart}, G.~P., \& {Meiring}, J.~G. 2006, in \procspie, Vol. 6267, Society of Photo-Optical Instrumentation Engineers (SPIE) Conference Series, 62670Z, \dodoi{10.1117/12.673750}

\bibitem[{{Burgh} {et~al.}(2003){Burgh}, {Nordsieck}, {Kobulnicky}, {Williams}, {O'Donoghue}, {Smith}, \& {Percival}}]{burgh2003}
{Burgh}, E.~B., {Nordsieck}, K.~H., {Kobulnicky}, H.~A., {et~al.} 2003, in Society of Photo-Optical Instrumentation Engineers (SPIE) Conference Series, Vol. 4841, Instrument Design and Performance for Optical/Infrared Ground-based Telescopes, ed. M.~{Iye} \& A.~F.~M. {Moorwood}, 1463--1471, \dodoi{10.1117/12.460312}

\bibitem[{{Catala} {et~al.}(2013){Catala}, {Crawford}, {Buckley}, {Pickering}, {Wilson}, {Butterley}, {Shepherd}, {Marang}, {Matshaya}, \& {Fourie}}]{catala2013}
{Catala}, L., {Crawford}, S.~M., {Buckley}, D.~A.~H., {et~al.} 2013, \mnras, 436, 590, \dodoi{10.1093/mnras/stt1602}

\bibitem[{{Courtois} {et~al.}(2019){Courtois}, {Kraan-Korteweg}, {Dupuy}, {Graziani}, \& {Libeskind}}]{courtois2019}
{Courtois}, H.~M., {Kraan-Korteweg}, R.~C., {Dupuy}, A., {Graziani}, R., \& {Libeskind}, N.~I. 2019, \mnras, 490, L57, \dodoi{10.1093/mnrasl/slz146}

\bibitem[{{Cuesta-Bolao} \& {Serna}(2003)}]{Cuesta-Bolao2003}
{Cuesta-Bolao}, M.~J., \& {Serna}, A. 2003, \aap, 405, 917, \dodoi{10.1051/0004-6361:20030567}

\bibitem[{{De Propris}(2017)}]{depropris2017}
{De Propris}, R. 2017, \mnras, 465, 4035, \dodoi{10.1093/mnras/stw2980}

\bibitem[{{De Propris} \& {Christlein}(2009)}]{depropris2009}
{De Propris}, R., \& {Christlein}, D. 2009, Astronomische Nachrichten, 330, 943, \dodoi{10.1002/asna.200911268}

\bibitem[{{de Propris} {et~al.}(1998){de Propris}, {Eisenhardt}, {Stanford}, \& {Dickinson}}]{depropris1998}
{de Propris}, R., {Eisenhardt}, P.~R., {Stanford}, S.~A., \& {Dickinson}, M. 1998, \apjl, 503, L45, \dodoi{10.1086/311531}

\bibitem[{{Devereux} {et~al.}(2009){Devereux}, {Hriljac}, {Willner}, {Ashby}, \& {Willmer}}]{devereux2009}
{Devereux}, N., {Hriljac}, P., {Willner}, S.~P., {Ashby}, M.~L.~N., \& {Willmer}, C.~N.~A. 2009, in Astronomical Society of the Pacific Conference Series, Vol. 419, Galaxy Evolution: Emerging Insights and Future Challenges, ed. S.~{Jogee}, I.~{Marinova}, L.~{Hao}, \& G.~A. {Blanc}, 171.
\newblock \doarXiv{0902.0778}

\bibitem[{{Devereux} {et~al.}(1987){Devereux}, {Becklin}, \& {Scoville}}]{devereux1987}
{Devereux}, N.~A., {Becklin}, E.~E., \& {Scoville}, N. 1987, \apj, 312, 529, \dodoi{10.1086/164899}

\bibitem[{{Dressler}(1980)}]{dressler1980}
{Dressler}, A. 1980, \apj, 236, 351, \dodoi{10.1086/157753}

\bibitem[{{Gavazzi} {et~al.}(1996){Gavazzi}, {Pierini}, \& {Boselli}}]{gavazzi1996}
{Gavazzi}, G., {Pierini}, D., \& {Boselli}, A. 1996, \aap, 312, 397

\bibitem[{{Gladders} {et~al.}(1998){Gladders}, {L{\'o}pez-Cruz}, {Yee}, \& {Kodama}}]{gladders1998}
{Gladders}, M.~D., {L{\'o}pez-Cruz}, O., {Yee}, H.~K.~C., \& {Kodama}, T. 1998, \apj, 501, 571, \dodoi{10.1086/305858}

\bibitem[{{Gladders} \& {Yee}(2000)}]{Gladders2000}
{Gladders}, M.~D., \& {Yee}, H.~K.~C. 2000, \aj, 120, 2148, \dodoi{10.1086/301557}

\bibitem[{{Hatamkhani} {et~al.}(2023){Hatamkhani}, {Kraan-Korteweg}, {Blyth}, {Said}, \& {Elagali}}]{hatamkhani2023}
{Hatamkhani}, N., {Kraan-Korteweg}, R.~C., {Blyth}, S.~L., {Said}, K., \& {Elagali}, A. 2023, \mnras, 522, 2223, \dodoi{10.1093/mnras/stad1134}

\bibitem[{{Healy} {et~al.}(2021){Healy}, {Blyth}, {Verheijen}, {Hess}, {Serra}, {van der Hulst}, {Jarrett}, {Yim}, \& {J{\'o}zsa}}]{healy2021c}
{Healy}, J., {Blyth}, S.~L., {Verheijen}, M.~A.~W., {et~al.} 2021, \aap, 650, A76, \dodoi{10.1051/0004-6361/202038738}

\bibitem[{{Henriques} {et~al.}(2015){Henriques}, {White}, {Thomas}, {Angulo}, {Guo}, {Lemson}, {Springel}, \& {Overzier}}]{henriques2015}
{Henriques}, B. M.~B., {White}, S. D.~M., {Thomas}, P.~A., {et~al.} 2015, \mnras, 451, 2663, \dodoi{10.1093/mnras/stv705}

\bibitem[{{Hilton} {et~al.}(2018){Hilton}, {Hasselfield}, {Sif{\'o}n}, {Battaglia}, {Aiola}, {Bharadwaj}, {Bond}, {Choi}, {Crichton}, {Datta}, {Devlin}, {Dunkley}, {D{\"u}nner}, {Gallardo}, {Gralla}, {Hincks}, {Ho}, {Hubmayr}, {Huffenberger}, {Hughes}, {Koopman}, {Kosowsky}, {Louis}, {Madhavacheril}, {Marriage}, {Maurin}, {McMahon}, {Miyatake}, {Moodley}, {N{\ae}ss}, {Nati}, {Newburgh}, {Niemack}, {Oguri}, {Page}, {Partridge}, {Schmitt}, {Sievers}, {Spergel}, {Staggs}, {Trac}, {van Engelen}, {Vavagiakis}, \& {Wollack}}]{hilton2018}
{Hilton}, M., {Hasselfield}, M., {Sif{\'o}n}, C., {et~al.} 2018, \apjs, 235, 20, \dodoi{10.3847/1538-4365/aaa6cb}

\bibitem[{{Huchra} {et~al.}(2012){Huchra}, {Macri}, {Masters}, {Jarrett}, {Berlind}, {Calkins}, {Crook}, {Cutri}, {Erdo{\v g}du}, {Falco}, {George}, {Hutcheson}, {Lahav}, {Mader}, {Mink}, {Martimbeau}, {Schneider}, {Skrutskie}, {Tokarz}, \& {Westover}}]{huchra2012}
{Huchra}, J.~P., {Macri}, L.~M., {Masters}, K.~L., {et~al.} 2012, \apjs, 199, 26, \dodoi{10.1088/0067-0049/199/2/26}

\bibitem[{{Jarrett}(2000)}]{jarrett2000b-k}
{Jarrett}, T.~H. 2000, \pasp, 112, 1008, \dodoi{10.1086/316603}

\bibitem[{{Jarrett} {et~al.}(2000){Jarrett}, {Chester}, {Cutri}, {Schneider}, {Skrutskie}, \& {Huchra}}]{jarrett2000A}
{Jarrett}, T.~H., {Chester}, T., {Cutri}, R., {et~al.} 2000, \aj, 119, 2498, \dodoi{10.1086/301330}

\bibitem[{{Jerjen} \& {Tammann}(1997)}]{jerjen1997}
{Jerjen}, H., \& {Tammann}, G.~A. 1997, \aap, 321, 713

\bibitem[{{Jones} {et~al.}(2006){Jones}, {Peterson}, {Colless}, \& {Saunders}}]{jones2006}
{Jones}, D.~H., {Peterson}, B.~A., {Colless}, M., \& {Saunders}, W. 2006, \mnras, 369, 25, \dodoi{10.1111/j.1365-2966.2006.10291.x}

\bibitem[{{Kirby} {et~al.}(2008){Kirby}, {Jerjen}, {Ryder}, \& {Driver}}]{kirby2008}
{Kirby}, E.~M., {Jerjen}, H., {Ryder}, S.~D., \& {Driver}, S.~P. 2008, \aj, 136, 1866, \dodoi{10.1088/0004-6256/136/5/1866}

\bibitem[{{Kocevski} \& {Ebeling}(2006)}]{kocevski2006}
{Kocevski}, D.~D., \& {Ebeling}, H. 2006, \apj, 645, 1043, \dodoi{10.1086/503666}

\bibitem[{{Kraan-Korteweg}(2000)}]{kraan2000catalog}
{Kraan-Korteweg}, R.~C. 2000, \aaps, 141, 123, \dodoi{10.1051/aas:2000314}

\bibitem[{{Kraan-Korteweg} {et~al.}(2017){Kraan-Korteweg}, {Cluver}, {Bilicki}, {Jarrett}, {Colless}, {Elagali}, {B{\"o}hringer}, \& {Chon}}]{kraan2017}
{Kraan-Korteweg}, R.~C., {Cluver}, M.~E., {Bilicki}, M., {et~al.} 2017, \mnras, 466, L29, \dodoi{10.1093/mnrasl/slw229}

\bibitem[{{Kurtz} \& {Mink}(1998)}]{kurtz1998}
{Kurtz}, M.~J., \& {Mink}, D.~J. 1998, \pasp, 110, 934, \dodoi{10.1086/316207}

\bibitem[{{Lee} {et~al.}(2016){Lee}, {Rey}, {Hilker}, {Sheen}, \& {Yi}}]{lee2016}
{Lee}, Y., {Rey}, S.-C., {Hilker}, M., {Sheen}, Y.-K., \& {Yi}, S.~K. 2016, \apj, 822, 92, \dodoi{10.3847/0004-637X/822/2/92}

\bibitem[{{Lin} {et~al.}(2004){Lin}, {Mohr}, \& {Stanford}}]{lin2004}
{Lin}, Y.-T., {Mohr}, J.~J., \& {Stanford}, S.~A. 2004, in Bulletin of the American Astronomical Society, Vol.~36, American Astronomical Society Meeting Abstracts \#204, 731

\bibitem[{{L{\'o}pez-Cruz} {et~al.}(2004){L{\'o}pez-Cruz}, {Barkhouse}, \& {Yee}}]{lopez2004}
{L{\'o}pez-Cruz}, O., {Barkhouse}, W.~A., \& {Yee}, H.~K.~C. 2004, \apj, 614, 679, \dodoi{10.1086/423664}

\bibitem[{{Marinoni} {et~al.}(1999){Marinoni}, {Monaco}, {Giuricin}, \& {Costantini}}]{marinoni1999}
{Marinoni}, C., {Monaco}, P., {Giuricin}, G., \& {Costantini}, B. 1999, \apj, 521, 50, \dodoi{10.1086/307516}

\bibitem[{{McDonald} {et~al.}(2009){McDonald}, {Courteau}, \& {Tully}}]{mcdonald2009}
{McDonald}, M., {Courteau}, S., \& {Tully}, R.~B. 2009, \mnras, 394, 2022, \dodoi{10.1111/j.1365-2966.2009.14442.x}

\bibitem[{{McNaught-Roberts} {et~al.}(2014){McNaught-Roberts}, {Norberg}, {Baugh}, {Lacey}, {Loveday}, {Peacock}, {Baldry}, {Bland-Hawthorn}, {Brough}, {Driver}, {Robotham}, \& {V{\'a}zquez-Mata}}]{McNaught2014}
{McNaught-Roberts}, T., {Norberg}, P., {Baugh}, C., {et~al.} 2014, \mnras, 445, 2125, \dodoi{10.1093/mnras/stu1886}

\bibitem[{{Mei} {et~al.}(2006{\natexlab{a}}){Mei}, {Blakeslee}, {Stanford}, {Holden}, {Rosati}, {Strazzullo}, {Homeier}, {Postman}, {Franx}, {Rettura}, {Ford}, {Illingworth}, {Ettori}, {Bouwens}, {Demarco}, {Martel}, {Clampin}, {Hartig}, {Eisenhardt}, {Ardila}, {Bartko}, {Ben{\'\i}tez}, {Bradley}, {Broadhurst}, {Brown}, {Burrows}, {Cheng}, {Cross}, {Feldman}, {Golimowski}, {Goto}, {Gronwall}, {Infante}, {Kimble}, {Krist}, {Lesser}, {Menanteau}, {Meurer}, {Miley}, {Motta}, {Sirianni}, {Sparks}, {Tran}, {Tsvetanov}, {White}, \& {Zheng}}]{mei2006a}
{Mei}, S., {Blakeslee}, J.~P., {Stanford}, S.~A., {et~al.} 2006{\natexlab{a}}, \apj, 639, 81, \dodoi{10.1086/499259}

\bibitem[{{Mei} {et~al.}(2006{\natexlab{b}}){Mei}, {Holden}, {Blakeslee}, {Rosati}, {Postman}, {Jee}, {Rettura}, {Sirianni}, {Demarco}, {Ford}, {Franx}, {Homeier}, \& {Illingworth}}]{mei2006b}
{Mei}, S., {Holden}, B.~P., {Blakeslee}, J.~P., {et~al.} 2006{\natexlab{b}}, \apj, 644, 759, \dodoi{10.1086/503826}

\bibitem[{{Merluzzi} {et~al.}(2010){Merluzzi}, {Mercurio}, {Haines}, {Smith}, {Busarello}, \& {Lucey}}]{merluzzi2010}
{Merluzzi}, P., {Mercurio}, A., {Haines}, C.~P., {et~al.} 2010, \mnras, 402, 753, \dodoi{10.1111/j.1365-2966.2009.15929.x}

\bibitem[{{Mitchell} {et~al.}(2018){Mitchell}, {Lacey}, {Lagos}, {Frenk}, {Bower}, {Cole}, {Helly}, {Schaller}, {Gonzalez-Perez}, \& {Theuns}}]{mitchell2018}
{Mitchell}, P.~D., {Lacey}, C.~G., {Lagos}, C. D.~P., {et~al.} 2018, \mnras, 474, 492, \dodoi{10.1093/mnras/stx2770}

\bibitem[{{Mobasher} {et~al.}(2003){Mobasher}, {Colless}, {Carter}, {Poggianti}, {Bridges}, {Kranz}, {Komiyama}, {Kashikawa}, {Yagi}, \& {Okamura}}]{mobasher2003}
{Mobasher}, B., {Colless}, M., {Carter}, D., {et~al.} 2003, \apj, 587, 605, \dodoi{10.1086/368305}

\bibitem[{{Monet}(1998)}]{monet1998}
{Monet}, D.~G. 1998, in American Astronomical Society Meeting Abstracts, Vol. 193, American Astronomical Society Meeting Abstracts, 120.03

\bibitem[{{Paolillo} {et~al.}(2001){Paolillo}, {Andreon}, {Longo}, {Puddu}, {Gal}, {Scaramella}, {Djorgovski}, \& {de Carvalho}}]{paolillo2001}
{Paolillo}, M., {Andreon}, S., {Longo}, G., {et~al.} 2001, \aap, 367, 59, \dodoi{10.1051/0004-6361:20000442}

\bibitem[{{Papovich} {et~al.}(2010){Papovich}, {Momcheva}, {Willmer}, {Finkelstein}, {Finkelstein}, {Tran}, {Brodwin}, {Dunlop}, {Farrah}, {Khan}, {Lotz}, {McCarthy}, {McLure}, {Rieke}, {Rudnick}, {Sivanandam}, {Pacaud}, \& {Pierre}}]{papovich2010}
{Papovich}, C., {Momcheva}, I., {Willmer}, C.~N.~A., {et~al.} 2010, \apj, 716, 1503, \dodoi{10.1088/0004-637X/716/2/1503}

\bibitem[{{Ramatsoku} {et~al.}(2020){Ramatsoku}, {Verheijen}, {Kraan-Korteweg}, {Jarrett}, {Said}, \& {Schr{\"o}der}}]{ramatsoku2020}
{Ramatsoku}, M., {Verheijen}, M.~A.~W., {Kraan-Korteweg}, R.~C., {et~al.} 2020, \aap, 644, A107, \dodoi{10.1051/0004-6361/202038342}

\bibitem[{{Rodrigues} {et~al.}(2017){Rodrigues}, {Vernon}, \& {Bower}}]{rodrig2017}
{Rodrigues}, L. F.~S., {Vernon}, I., \& {Bower}, R.~G. 2017, \mnras, 466, 2418, \dodoi{10.1093/mnras/stw3269}

\bibitem[{{Rykoff} {et~al.}(2016){Rykoff}, {Rozo}, {Hollowood}, {Bermeo-Hernandez}, {Jeltema}, {Mayers}, {Romer}, {Rooney}, {Saro}, {Vergara Cervantes}, {Wechsler}, {Wilcox}, {Abbott}, {Abdalla}, {Allam}, {Annis}, {Benoit-L{\'e}vy}, {Bernstein}, {Bertin}, {Brooks}, {Burke}, {Capozzi}, {Carnero Rosell}, {Carrasco Kind}, {Castander}, {Childress}, {Collins}, {Cunha}, {D'Andrea}, {da Costa}, {Davis}, {Desai}, {Diehl}, {Dietrich}, {Doel}, {Evrard}, {Finley}, {Flaugher}, {Fosalba}, {Frieman}, {Glazebrook}, {Goldstein}, {Gruen}, {Gruendl}, {Gutierrez}, {Hilton}, {Honscheid}, {Hoyle}, {James}, {Kay}, {Kuehn}, {Kuropatkin}, {Lahav}, {Lewis}, {Lidman}, {Lima}, {Maia}, {Mann}, {Marshall}, {Martini}, {Melchior}, {Miller}, {Miquel}, {Mohr}, {Nichol}, {Nord}, {Ogando}, {Plazas}, {Reil}, {Sahl{\'e}n}, {Sanchez}, {Santiago}, {Scarpine}, {Schubnell}, {Sevilla-Noarbe}, {Smith}, {Soares-Santos}, {Sobreira}, {Stott}, {Suchyta}, {Swanson}, {Tarle}, {Thomas}, {Tucker}, {Uddin}, {Viana}, {Vikram}, {Walker}, {Zhang}, \& {DES
  Collaboration}}]{rykoff2016}
{Rykoff}, E.~S., {Rozo}, E., {Hollowood}, D., {et~al.} 2016, \apjs, 224, 1, \dodoi{10.3847/0067-0049/224/1/1}

\bibitem[{{Said} {et~al.}(2015){Said}, {Kraan-Korteweg}, \& {Jarrett}}]{said2015}
{Said}, K., {Kraan-Korteweg}, R.~C., \& {Jarrett}, T.~H. 2015, \mnras, 447, 1618, \dodoi{10.1093/mnras/stu2496}

\bibitem[{{Schechter}(1976)}]{schechter1976}
{Schechter}, P. 1976, \apj, 203, 297, \dodoi{10.1086/154079}

\bibitem[{{Scrimgeour} {et~al.}(2016){Scrimgeour}, {Davis}, {Blake}, {Staveley-Smith}, {Magoulas}, {Springob}, {Beutler}, {Colless}, {Johnson}, {Jones}, {Koda}, {Lucey}, {Ma}, {Mould}, \& {Poole}}]{scrimgeour2016}
{Scrimgeour}, M.~I., {Davis}, T.~M., {Blake}, C., {et~al.} 2016, \mnras, 455, 386, \dodoi{10.1093/mnras/stv2146}

\bibitem[{{Skelton} {et~al.}(2009){Skelton}, {Woudt}, \& {Kraan-Korteweg}}]{skelton2009}
{Skelton}, R.~E., {Woudt}, P.~A., \& {Kraan-Korteweg}, R.~C. 2009, \mnras, 396, 2367, \dodoi{10.1111/j.1365-2966.2009.14905.x}

\bibitem[{{Skrutskie} {et~al.}(2006){Skrutskie}, {Cutri}, {Stiening}, {Weinberg}, {Schneider}, {Carpenter}, {Beichman}, {Capps}, {Chester}, {Elias}, {Huchra}, {Liebert}, {Lonsdale}, {Monet}, {Price}, {Seitzer}, {Jarrett}, {Kirkpatrick}, {Gizis}, {Howard}, {Evans}, {Fowler}, {Fullmer}, {Hurt}, {Light}, {Kopan}, {Marsh}, {McCallon}, {Tam}, {Van Dyk}, \& {Wheelock}}]{skrutskie2006}
{Skrutskie}, M.~F., {Cutri}, R.~M., {Stiening}, R., {et~al.} 2006, \aj, 131, 1163, \dodoi{10.1086/498708}

\bibitem[{{Snyder} {et~al.}(2012){Snyder}, {Brodwin}, {Mancone}, {Zeimann}, {Stanford}, {Gonzalez}, {Stern}, {Eisenhardt}, {Brown}, {Dey}, {Jannuzi}, \& {Perlmutter}}]{snyder2012}
{Snyder}, G.~F., {Brodwin}, M., {Mancone}, C.~M., {et~al.} 2012, \apj, 756, 114, \dodoi{10.1088/0004-637X/756/2/114}

\bibitem[{{Sorce} {et~al.}(2017){Sorce}, {Colless}, {Kraan-Korteweg}, \& {Gottl{\"o}ber}}]{sorce2017}
{Sorce}, J.~G., {Colless}, M., {Kraan-Korteweg}, R.~C., \& {Gottl{\"o}ber}, S. 2017, \mnras, 471, 3087, \dodoi{10.1093/mnras/stx1800}

\bibitem[{{Springob} {et~al.}(2014){Springob}, {Magoulas}, {Colless}, {Mould}, {Erdo{\u g}du}, {Jones}, {Lucey}, {Campbell}, \& {Fluke}}]{springob2014}
{Springob}, C.~M., {Magoulas}, C., {Colless}, M., {et~al.} 2014, \mnras, 445, 2677, \dodoi{10.1093/mnras/stu1743}

\bibitem[{{Springob} {et~al.}(2016){Springob}, {Hong}, {Staveley-Smith}, {Masters}, {Macri}, {Koribalski}, {Jones}, {Jarrett}, {Magoulas}, \& {Erdo{\u{g}}du}}]{springob2016}
{Springob}, C.~M., {Hong}, T., {Staveley-Smith}, L., {et~al.} 2016, \mnras, 456, 1886, \dodoi{10.1093/mnras/stv2648}

\bibitem[{{Stanford} {et~al.}(1998){Stanford}, {Eisenhardt}, \& {Dickinson}}]{stanford1998}
{Stanford}, S.~A., {Eisenhardt}, P.~R., \& {Dickinson}, M. 1998, \apj, 492, 461, \dodoi{10.1086/305050}

\bibitem[{{Stott} {et~al.}(2009){Stott}, {Pimbblet}, {Edge}, {Smith}, \& {Wardlow}}]{stott2009}
{Stott}, J.~P., {Pimbblet}, K.~A., {Edge}, A.~C., {Smith}, G.~P., \& {Wardlow}, J.~L. 2009, \mnras, 394, 2098, \dodoi{10.1111/j.1365-2966.2009.14477.x}

\bibitem[{{Strazzullo} {et~al.}(2016){Strazzullo}, {Daddi}, {Gobat}, {Valentino}, {Pannella}, {Dickinson}, {Renzini}, {Brammer}, {Onodera}, {Finoguenov}, {Cimatti}, {Carollo}, \& {Arimoto}}]{strazzullo2016A}
{Strazzullo}, V., {Daddi}, E., {Gobat}, R., {et~al.} 2016, \apjl, 833, L20, \dodoi{10.3847/2041-8213/833/2/L20}

\bibitem[{{Two-Micron All Sky Survey Science Team}(2020)}]{irsa97}
{Two-Micron All Sky Survey Science Team}. 2020, 2MASS All-Sky Extended Source Catalog,  IPAC, \dodoi{10.26131/IRSA97}

\bibitem[{{Woudt} {et~al.}(2008){Woudt}, {Kraan-Korteweg}, {Lucey}, {Fairall}, \& {Moore}}]{woudt2008}
{Woudt}, P.~A., {Kraan-Korteweg}, R.~C., {Lucey}, J., {Fairall}, A.~P., \& {Moore}, S.~A.~W. 2008, \mnras, 383, 445, \dodoi{10.1111/j.1365-2966.2007.12571.x}

\bibitem[{{Yahil} \& {Vidal}(1977)}]{yahil1977}
{Yahil}, A., \& {Vidal}, N.~V. 1977, \apj, 214, 347, \dodoi{10.1086/155257}

\bibitem[{{Yee} {et~al.}(1999){Yee}, {Gladders}, \& {L{\'o}pez-Cruz}}]{Yee1999A}
{Yee}, H.~K.~C., {Gladders}, M.~D., \& {L{\'o}pez-Cruz}, O. 1999, in Astronomical Society of the Pacific Conference Series, Vol. 191, Photometric Redshifts and the Detection of High Redshift Galaxies, ed. R.~{Weymann}, L.~{Storrie-Lombardi}, M.~{Sawicki}, \& R.~{Brunner}, 166

\end{thebibliography}
\bibliographystyle{aasjournal}



\end{document}